\documentclass[12pt,preprint]{aastex}

\usepackage{epsf}

\shorttitle{LBGs and the Ly$\alpha$ forest}
\shortauthors{Kollmeier et al.}

\newcommand{\lya}{Ly$\alpha$}
\newcommand{\kms}{\,{\rm km}\;{\rm s}^{-1}}

\newcommand{\hubunits}{\,\kms\;{\rm Mpc}^{-1}}

\newcommand{\hmpc}{\,h^{-1}\;{\rm Mpc}}
\newcommand{\hkpc}{\,h^{-1}\;{\rm kpc}}
\newcommand{\msun}{M_\odot}

\newcommand{\K}{\,{\rm K}}
\newcommand{\cm}{{\rm cm}}

\newcommand{\be}{\begin{equation}}
\newcommand{\ee}{\end{equation}}

\newcommand{\bb}{{\rm L}$11${\rm n}$128$ }
\newcommand{\ba}{{\rm L}$11${\rm n}$64$ }
\newcommand{\bc}{{\rm L}$22${\rm n}$128$ }
\newcommand{\bd}{{\rm L}$50${\rm n}$144$ }
\newcommand{\dunits}{{\,{\rm arcmin}^{-2}\, \Delta z^{-1}}}
\newcommand{\meand}{{\langle D \rangle}}
\newcommand{\msph}{{m_{\rm SPH}}}
\newcommand{\fsat}{{f_{\rm sat}}}
\newcommand{\ftrans}{{f_{\rm trans}}}
\newcommand{\tmax}{{\theta_{\rm max}}}
\newcommand{\teq}{t_{\rm eq}}
\newcommand{\tactive}{t_{\rm act}}
\newcommand{\fq}{f_Q}
\newcommand{\fgamma}{F_\Gamma}
\newcommand{\Gammaq}{\Gamma_{Q}}
\newcommand{\meands}{D_S}
\newcommand{\rmax}{{r_{\rm max}}}
\newcommand{\meanl}{{\langle L \rangle}}
\newcommand{\meann}{{\bar{N}}}
\newcommand{\dgal}{{\delta_{\rm gal}}}
\newcommand{\ddark}{{\delta_{\rm dm}}}
\newcommand{\siggal}{{\sigma_{\rm gal}}}
\newcommand{\sigdark}{{\sigma_{\rm dm}}}
\newcommand{\sigdm}{{\sigma_{\rm dm}}}
\newcommand{\sigD}{{\sigma_D}}
\newcommand{\rgaldm}{r_{\rm gal,dm}}
\newcommand{\rgalD}{r_{\rm gal,D}}
\newcommand{\rdarkD}{r_{\rm dm,D}}

\slugcomment{Accepted by ApJ}
\begin{document}
\title{Lyman Break Galaxies and the Lyman-alpha Forest}
\author{Juna A. Kollmeier$^1$, David H. Weinberg$^1$, Romeel Dav\'e,$^2$,
Neal Katz$^3$}

\footnotetext[1]
{Ohio State University, Dept.\ of Astronomy, Columbus, OH 43210,
jak,dhw@astronomy.ohio-state.edu}
\footnotetext[2]
{University of Arizona, Dept.\ of Astronomy, Tucson, AZ 85721,
rad@as.arizona.edu}
\footnotetext[3]
{University of Massachusetts, Dept.\ of Physics and Astronomy, Amherst,
MA 91003, nsk@kaka.phast.umass.edu}

\begin{abstract}
We use hydrodynamic cosmological simulations to predict correlations
between \lya\ forest absorption and the galaxy distribution at
redshift $z\approx 3$.  The probability distribution function (PDF) of
\lya\ flux decrements shifts systematically towards higher values in
the vicinity of galaxies, reflecting the overdense environments in
which these galaxies reside.  The predicted signal remains strong in
spectra smoothed over $50-200\kms$, allowing tests with moderate
resolution quasar spectra.  The strong bias of high redshift galaxies
towards high density regions imprints a clear signature on the flux
PDF, but the predictions are not sensitive to galaxy baryon mass or
star formation rate, and they are similar for galaxies and for dark
matter halos.  The dependence of the flux PDF on galaxy proximity is
sensitive to redshift determination errors, with rms errors of
$150-300 \kms$ substantially weakening the predicted trends.  On
larger scales, the mean galaxy overdensity in a cube of 5 or $10\hmpc$
(comoving) is strongly correlated with the mean \lya\ flux decrement
on a line of sight through the cube center.  The slope of the
correlation is $\sim 3$ times steeper for galaxies than for dark
matter as a result of galaxy bias.  The predicted large scale
correlation is in qualitative agreement with recently reported
observational results.  However, observations also show a drop in the
average absorption in the immediate vicinity of galaxies, which our
models do not predict even if we allow the galaxies or AGNs within
them to be ionizing sources.  This decreased absorption could be a
signature of galaxy feedback on the surrounding intergalactic medium,
perhaps via galactic winds.  We find that a simplified ``wind'' model
that eliminates neutral hydrogen in spheres around the galaxies can
marginally explain the data.  However, because peculiar velocities allow
gas at large distances to produce saturated absorption at the galaxy
redshift, these winds (or any other feedback mechanism) must extend to
comoving radii of $\sim 1.5 \hmpc$ to reproduce the observations.  We
also discuss the possibility that extended \lya\ emission from the
target galaxies ``fills in'' the expected \lya\ forest absorption at
small angular separations.
\end{abstract}

\section{Introduction}
\label{sec:intro}

The strong clustering of Lyman-break galaxies (LBGs) at $z\approx 3$,
comparable to that of present-day, optically selected galaxies, suggests
that they are highly biased tracers of the underlying dark matter
distribution \citep{adelberger98,adelberger02}.  This bias appears
to arise naturally in semi-analytic models and hydrodynamic
numerical simulations \citep[e.g.,][]{baugh98,governato98,katz99,
kauffmann99,cen00,benson01,pearce01,yoshikawa01,weinberg02a},
which predict that the luminous
members of the high redshift galaxy population reside in massive halos,
which in turn reside in regions of high background density
\citep{kaiser84,mof96,mow96}.  However, the bias of Lyman-break galaxies
is inferred by comparing their observed clustering to the predicted
clustering of dark matter, which depends on the assumed cosmological
model.  The \lya\ forest offers a tracer of structure
whose relation to the underlying
dark matter distribution appears to be well understood on theoretical
grounds, probing the same redshifts as the LBG population.
Correlations between LBGs and \lya\ forest absorption
therefore offer a natural and potentially powerful probe of the
relation between high redshift galaxies and the dark matter distribution,
and perhaps for the influence of high redshift galaxies on the
surrounding intergalactic medium (IGM).

In this paper we use smoothed particle hydrodynamics (SPH) simulations
to predict the correlations between $z=3$ galaxies and \lya\ forest
absorption.  Hydrodynamic simulations are ideal for this purpose,
since they simultaneously predict the locations and properties of
the galaxies \citep[e.g.][]{katz99,nagamine01,weinberg02b} and the
structure and ionization state of the intergalactic gas that produces
the \lya\ forest \citep[e.g.][]{cen94,zhang95,hernquist96,theuns98}.
However, the basic expectations can be understood in simple terms,
using the ``Fluctuating Gunn-Peterson Approximation''
\citep{bi97,croft97,rauch97,weinberg98,croft98}, which provides a
fairly accurate description of the \lya\ forest results from
full hydrodynamic calculations.  In this approximation, the \lya\
flux decrement at wavelength $\lambda$ is
related to the overdensity at a redshift
$z=\lambda/\lambda_{\alpha,{\rm rest}}-1$ along the line of sight by
\begin{equation}
D ~=~ 1 - \frac{F}{F_c} = 1-e^{-\tau} ~=~
  1 - \exp\left[ -A (\rho/\bar{\rho})^\beta\right] ~,
\label{eqn:fgpa}
\end{equation}
with
\begin{equation}
A = 0.694 \left(\frac{1+z}{4.0}\right)^6
          \left(\frac{\Omega_b h^2}{0.02}\right)^2
          \left(\frac{T_0}{6000\K}\right)^{-0.7}
          \left(\frac{h}{0.65}\right)^{-1}
          \left(\frac{H(z)/H_0}{5.12}\right)^{-1}
          \left(\frac{\Gamma}{1.5\times10^{12}\;\sec^{-1}}\right)^{-1}
	  ~.
\end{equation}
Here $h\equiv H_0/100\hubunits$, $\Omega_b$ is the baryon density parameter,
and $\Gamma$ is the HI photoionization rate due to the
diffuse UV background at redshift $z$.
This approximation assumes that all gas obeys a temperature-density relation
$T=T_0(\rho_b/\bar{\rho}_b)^\alpha$, which emerges from the
interplay between photoionization heating and adiabatic cooling
caused by cosmic expansion.  For typical reionization histories,
one expects $T_0 \sim 5000-20,000\K$ and
$\beta = 2-0.7\alpha \approx 1.6-1.8$ at $z\sim 3$, with
the higher $T_0$ and $\beta$ values arising if helium reionization
occurs close to this redshift \citep{hui97}.
Strictly speaking, the overdensity in equation~(\ref{eqn:fgpa})
is the gas overdensity, but in the moderate density regions
that contribute most of the \lya\ forest opacity, the gas pressure
is low, and the gas and dark matter trace each other fairly well.
This approximation ignores the effects of collisional ionization and
peculiar velocities.

 From equation~(\ref{eqn:fgpa}), one can see that the probability
distribution function (PDF) of \lya\ forest flux decrements,
$P(D)$, is closely related to the PDF of the underlying
density field, $P(\rho/\bar{\rho})$, with an effective smoothing
scale determined by a combination of pressure support and thermal
broadening along the line of sight.  The flux decrement PDF can
therefore provide a diagnostic for the amplitude of mass fluctuations
and for departures from Gaussianity in primordial fluctuations
\citep{weinberg99a,croft99,mcdonald00}.

The statistical measure that we focus on in this paper is the
{\it conditional} flux decrement PDF, the probability $P(D)$ that
a pixel in a \lya\ forest spectrum has a flux decrement in the
range $D\rightarrow D+dD$, conditioned on the presence of a galaxy
at the same redshift close to the line of sight.
(Since $D=1-F/F_c$, we will use the slightly less cumbersome
term ``conditional flux PDF,'' but we treat $D$ as the observable
of interest because it is an increasing function of density.)
This statistic can be measured by obtaining \lya\ forest spectra
in fields covered by LBG surveys.  The bias of galaxies towards
high density regions should reveal itself as a systematic shift
of $P(D)$ towards higher flux decrements in pixels close to galaxies,
with closer proximity yielding stronger shifts.  Conversely, feedback
of galaxies on the local IGM via ionization or galactic winds
might shift the PDF towards lower flux decrements at small separations.

A complete measurement of the conditional flux PDF will require ambitious,
time-consuming surveys on large telescopes, since each LBG-quasar pair
contributes only a single point to the distribution.  However, characteristics
of the conditional PDF like the mean decrement or the fraction of
saturated and ``transparent'' pixels can be measured with fewer
data points, and we present predictions for these quantities in addition
to the full PDF.  We also show that the signature of galaxy proximity
should persist in spectra smoothed over $50-200\kms$, allowing
theoretical predictions to be tested using relatively low resolution
quasar spectra. With higher resolution spectra the dependence of
the conditional PDF on spectral smoothing offers a further test of
the models.

K.\ Adelberger and collaborators have been carrying out a survey along
the lines envisioned here 
(\citealt{adelberger02}, hereafter ASSP;
\citealt{adelberger02b}, hereafter ASPS).
While we have not designed
our study primarily for comparison with their results, we will discuss
the comparison between our predictions and some
of their measurements in \S\ref{sec:discussion} below.  Inspired by
their work, we also consider an additional statistic, the mean galaxy
overdensity in $5\,h^{-1}$ and
$10\hmpc$ (comoving) cubes as a function of the mean
decrement on a line of sight through the middle of the cube.
This statistic characterizes the large scale correlations between
galaxies and IGM overdensity, while the conditional flux PDF
characterizes correlations on smaller scales.
McDonald, Miralda-Escud\'e, \& Cen (\citeyear{mcdonald02})
have presented predictions for this statistic using dark matter in
N-body simulations, and here we show predictions for the biased galaxy
population of a hydrodynamic simulation.

There is substantial overlap between our investigation and the recent paper
of \cite{croft02a}.
As discussed in \S\ref{sec:discussion}, our results generally agree
well with theirs where we examine similar quantities.
\cite{croft02a} devote considerably more attention to models that
incorporate galactic winds, which they add to their simulations using
the post-processing approach of \cite{aguirre01}.  Here we concentrate
on the direct predictions of our simulations, which incorporate
thermal feedback in the local ISM that is usually radiated away
before it can drive a galactic wind.  However, we do
estimate the possible effects of galaxy photoionization, and
of photoionization by recurrent AGN activity associated with galaxies,
and we briefly discuss the possible influence of stronger winds.
We describe our simulations, galaxy identification, and spectral
extraction in the next section.  We present our results for the
conditional flux PDF in \S\ref{sec:cpdf} and for the correlation
between mean decrement and galaxy counts in cubic cells in \S\ref{sec:cube}.
In \S\ref{sec:discussion}, we briefly review our results, then
discuss them in relation to the theoretical study of \cite{croft02a}
and the observational analyses of ASSP, concluding
with some remarks about future directions.

\section{Method}
\label{sec:method}
Our results are derived from SPH simulations of an inflationary cold
dark matter model with a cosmological constant ($\Lambda$CDM), with
cosmological parameters $\Omega_m=0.4$, $\Omega_\Lambda=0.6$,
$h=0.65$, $\Omega_b=0.02h^{-2}=0.0473$, an inflationary spectral
index $n=0.95$, and a power spectrum normalization $\sigma_8=0.80$.
These simulations use a parallel implementation of TreeSPH
(\citealt{hernquist89}; \citealt{katz96}, hereafter KWH;
\citealt{dave97}), incorporating radiative cooling and star
formation as described by KWH.  We rely mainly on two simulations,
one that uses $128^3$ SPH particles and $128^3$ dark matter particles
in a periodic cube $22.222\hmpc$ (comoving) on a side, and one that uses
$2\times 144^3$ particles in a cube $50\hmpc$ on a side.  We use
two additional simulations to investigate numerical resolution effects.
We refer to simulations with the nomenclature LXnY, where the box
size is X$\hmpc$ and the particle number is Y$^3$ for each species.
The main simulations are thus L22n128 and L50n144, and the additional
simulations are L11n64 and L11n128.  The properties of all of these
simulations are summarized in Table~\ref{tab:sims}.
L22n128 and L11n64 have the same resolution as simulations that
we have previously used to study the \lya\ forest
\citep{hernquist96,dave99}, while L11n128 has a mass resolution
a factor of eight higher.  The L50n144 simulation was designed primarily for
studies of galaxy assembly, galaxy clustering, and the low redshift
intergalactic medium \citep{dave01,dave02,croft01,chen02,fardal02,murali02,
weinberg02a}.  It has a substantially larger volume than our other
simulations, but its mass resolution is a factor of eight lower than
that of L22n128.  We analyze all of the simulations at redshift $z=3$,
close to the mean redshift of the main LBG samples.

With the exception of L50n144, all of the simulations were evolved with the
photoionizing UV background calculated by \citet{haardt96}.
At the processing stage, the intensity of the background is adjusted
by small factors ($0.75-1.16$, depending on the simulation)
so that the mean decrement of the \lya\ forest spectra extracted
from the simulation matches the value $D=0.36$ found by
\cite{press93} at $z=3$ (\citealt{bernardi03} find a nearly
identical value with a much larger quasar sample).
The L50n144 simulation was run with no ionizing background
(for the reasons discussed by \citealt{weinberg97a}),
so its unshocked gas cools adiabatically to unrealistically low
temperatures.  We therefore modify the temperatures of all gas particles
that have $(\rho/\bar{\rho})<100$ and
$T<10^{4}(\rho/\bar{\rho})^{0.6}\K$, setting them to
$T=10^{4}(\rho/\bar{\rho})^{0.6}\K$.
We do not modify the temperatures of hotter particles.
In this way, we mimic the results expected if we had used an ionizing
background during evolution, and the tests in \S\ref{sec:numerics}
below indicate that this approach is adequate for our purposes.
We again extract spectra with an ionizing background intensity
chosen to reproduce the \cite{press93} mean flux decrement.

The star formation algorithm is discussed in detail by KWH.
In brief, cold gas with physical density $n_H > 0.1\,\cm^{-3}$
and overdensity $\rho/\bar{\rho}>1000$ is
converted into stars on a timescale
determined by the local dynamical and cooling times.  Feedback is
deposited as thermal energy in the surrounding medium,
and it is usually radiated away before the interstellar medium
becomes hot enough to drive a ``galactic wind.''  The SKID
algorithm\footnote{{\tt http://www-hpcc.astro.washington.edu/tools/skid.html}}
(Spline-Kernel Interpolated DENMAX; see KWH and \citealt{gelb94})
is used to identify clumps of stars and cold gas
$(T<30,000\K,~\rho/\bar{\rho}>1000)$ that correspond to ``galaxies.''
Because these systems are extremely overdense, there is no ambiguity
in their identification, and their masses are robust to changes
in the details of the radiative cooling or star formation treatment
(KWH; \citealt{weinberg97a}).
We restrict our attention to galaxies with baryonic mass
(stars plus cold gas) $M_b \geq 64m_{\rm SPH}$, since our comparisons
among simulations of different resolution indicate that this
population is insensitive to resolution effects.

The simulations produce a population of high redshift galaxies whose
clustering, star formation properties, and sub-mm luminosities are
in approximate agreement with observations
\citep{katz99,weinberg99b,weinberg02a,weinberg02b,fardal02},
though the uncertainties in some of these comparisons are substantial.
Correlations with the \lya\ forest offer the possibility of testing
the simulation predictions for the relation between these galaxies
and the underlying mass distribution.  The simulated galaxies span
a wide dynamic range in baryonic mass and star formation rate
(SFR), and we want to know whether the predicted correlations with the
\lya\ forest depend substantially on galaxy properties.
Rather than model the galaxy spectral energy distributions in detail,
we define galaxy samples above thresholds in baryonic mass or SFR
and characterize them by their space density,
\begin{equation}
\Sigma=\frac{N_{\rm gal}}{\theta_{\rm box}^2 \times (\Delta z)_{\rm box}} ~,
\end{equation}
where $N_{\rm gal}$ is the number of galaxies above the threshold in
the simulation box,
$\theta_{\rm box}$ is the projected size of the box in
arc-minutes, and $(\Delta z)_{\rm box}$ is the redshift depth of the box.
The units of $\Sigma$ are thus number per square arc-minute
per unit redshift (which we will write $\dunits$),
and it is worth emphasizing that ``per unit redshift''
makes $\Sigma$ a space density rather than a surface density.
To the extent that galaxies' rest-frame UV luminosities are
monotonically related to their instantaneous star formation rates,
samples selected above an SFR threshold should correspond to
observed LBG samples above an optical
magnitude limit.  Typical LBG spectroscopic samples to $R\sim 25.5$
have $\Sigma \sim 1-3\dunits$ \citep{steidel96,adelberger98}.

\lya\ forest spectra are extracted using TIPSY\footnote{{\tt
http://www-hpcc.astro.washington.edu/tools/tipsy/tipsy.html} }, as
described by \cite{hernquist96}.  Briefly, the optical depth at each
position is calculated from the neutral hydrogen mass, neutral
mass-weighted velocity, and neutral mass-weighted temperature, derived
by integrating through all of the SPH smoothing kernels that overlap
that spatial position.  We extract 1200 spectra along random lines of
sight (400 in each projection) for each simulation.  We use ``pixels''
of $1.5-3\kms$, small enough that variations on the thermal broadening
scale $\sim 10\kms$ are fully resolved.  As discussed below, we
compute statistics after smoothing these spectra with Gaussians of
$1\sigma$ width $10-200\kms$, with the smallest of these scales
representing the approximate spectral resolution of typical high
resolution QSO spectra.  We do not add noise to our artificial
spectra; in general, noise will act to broaden measured distribution
functions but will average out in measurements of mean decrements.
To compute the conditional flux PDF statistics discussed in the next
section, we search around each sightline for nearby galaxies in the
cylindrical volume defined by a given search radius and the depth of
the box.  This method yields few matches at small angular separations
or for rare galaxy classes, so we supplement the random spectra with
spectra extracted close to the known positions of the galaxies, with
the distribution of projected separations that would be expected for
random positioning.

\section{The Conditional Flux PDF}
\label{sec:cpdf}

We begin by examining the
physical conditions of the IGM in the vicinity of galaxies.
Figure~\ref{fig:profiles}, based on the L22n128 simulation,
shows the median value, interquartile range, and 10--90\% range
of the temperature, gas overdensity, 
neutral hydrogen density, and neutral hydrogen fraction 
in pixels that lie at the same redshift as
a simulated galaxy at transverse separation $r_t$ (in comoving $\hmpc$).
For our adopted cosmological parameters, an angular separation of $1'$
at $z=3$ corresponds to $r_t=1.19\hmpc$ (comoving), and to
$Hr_t = 152\kms$.
As expected, the gas density increases close to galaxies.
Typical temperatures are also higher in the vicinity of galaxies.
In the median, this temperature increase is mostly accounted for
by the increased photoionization heating at higher densities, but
the spread to high temperatures is probably an effect of shock 
heating as gas falls into collapsed regions.  In our simulations,
supernova heating should have little impact at the separations
shown in Figure~\ref{fig:profiles}.
While higher temperatures counteract the
influence of higher densities, the density effect is stronger,
and typical {\it neutral} hydrogen densities increase sharply in
the neighborhood of galaxies.  The median neutral fraction is roughly
constant at $\sim 10^{-5}$, increasing at $r_t \leq 100\hkpc$.

Figure~\ref{fig:profiles} motivates our use of the statistic described
in the introduction, the conditional PDF of the flux decrement $D$ in
pixels within angular separation $\tmax$ of a galaxy at the same
redshift.  
We will investigate the dependence of this statistic on the angular
separation threshold $\tmax$, on the smoothing of the \lya\ forest
spectrum, on the velocity separation between the pixel redshift and
the galaxy redshift, and on the properties of the galaxies themselves.
In addition to the complete conditional PDF, we calculate three
quantities that summarize some of its properties: the (conditional)
mean decrement $\meand = \int_0^1 D P(D) dD$, the fraction $\fsat$ of
``saturated'' pixels with $D>0.9$, and the fraction $\ftrans$ of
``transparent'' pixels with $D<0.55$.  Before presenting our main
results from the L22n128 and L50n144 simulations, we first test the
sensitivity of the predictions to numerical parameters of the
simulations.

In general, one can consider the dependence of the flux PDF on both 
the angular separation $\Delta\theta$ between the galaxy and the
QSO sightline and the velocity separation $\Delta V$ between the
galaxy redshift and the pixel redshift.  For the most part, we concentrate
on the $\Delta V=0$ slice of this more general distribution
$P(D|\Delta\theta,\Delta V)$, so for each galaxy-sightline pair 
we consider the single pixel closest to the galaxy redshift, 
incorporating the effects of peculiar velocity on both the galaxy
redshift and the \lya\ forest spectrum.
Alternatively, one can define the conditional flux PDF in bins of
redshift-space separation 
$\Delta_r = \left[(d_A\Delta\theta)^2 + (\Delta V/H)^2\right]^{1/2}$,
where $d_A$ is the angular diameter distance and $H$ the Hubble
parameter at redshift $z$.  There are three disadvantages to the latter
approach: it requires a cosmological model to compute $d_A$ and $H(z)$
(which do not scale the same way with cosmological parameters),
peculiar velocities distort the relation between $\Delta V$ and
distance, and the statistical properties of the measured distributions
are more complex because each galaxy-quasar pair contributes correlated
values to $P(D)$ in different bins of $\Delta_r$.  However, the 
$\Delta_r$ definition has a significant advantage when computing the
conditional mean decrement from a small sample of moderate S/N spectra,
namely that each spectrum contributes many pixels and the noise 
averages out.  Our definition requires either high S/N per pixel or
a large number of galaxy-QSO pairs to obtain a precise measurement 
of the conditional mean decrement.  In practice, we recommend smoothing
spectra enough to ensure high S/N per pixel, since we show below that
smoothing over $\Delta V \sim 50-100\kms$ does not reduce the 
discriminating power of the measurements.  This way explicit smoothing
achieves the noise suppression that is obtained by binning in 
the $\Delta_r$ approach.  In \S\ref{sec:discussion}, we will 
calculate the mean decrement in bins of $\Delta_r$ for comparison
to Croft et al.'s (\citeyear{croft02a}) predictions and ASSP's measurements.

\subsection{Dependence on Numerical Parameters}
\label{sec:numerics}

We investigate the influence of mass resolution using the L11n128 and
L11n64 simulations, which have the same initial conditions and box
size but a factor of eight difference in particle mass.
Figure~\ref{fig:resolution} presents the conditional flux PDFs of
these simulations, in a format similar to that used in many of our
subsequent figures.  The left hand panel shows results for spectra
smoothed with a Gaussian of $1\sigma$ width $R_s=10\kms$,
and the right hand panel
for $R_s=50\kms$.  Solid curves show the unconditional PDF, computed
from all pixels in all of the randomly positioned spectra.
Dotted curves show the conditional PDF for pixels within $2'$ of
galaxies at the same redshift.  The physical mass threshold
for these galaxies is the same in the two simulations,
corresponding to $64\msph$ in L11n64 and to $512\msph$ in L11n128.
Dashed curves show the conditional flux PDF for angular separations
$\Delta\theta < 0.5'$.  Symbols along the top show the mean decrements
computed from these three PDFs.  Symbols along the right and
left hand vertical axes denote the fractions of saturated and
transparent pixels, respectively.

We will discuss the physics behind these conditional PDFs
in \S\ref{sec:results} below, though the basic behavior is clearly
what one would expect based upon the general discussion in \S\ref{sec:intro},
given the trends in Figure~\ref{fig:profiles}.
The crucial point of Figure~\ref{fig:resolution} is the similarity
between results represented by the bold lines/symbols and those represented
by the light lines/symbols, which come from the L11n128 and L11n64
simulations, respectively.  While there are systematic differences,
most notably in the number of transparent pixels for $\Delta\theta < 0.5'$,
these differences are quite small, especially compared to the
much stronger dependence on the separation itself.  Figure~\ref{fig:resolution}
implies that the L22n128 simulation, which has
the same mass resolution as L11n64 and a larger volume,
should yield reliable predictions for
galaxies above its $64\msph$ resolution threshold.

Comparing the
\ba and \bc simulations allows us to test the influence of box size
on the conditional flux PDF.  
The simulations cannot include density perturbations on scales larger than
the fundamental mode of the box, and this absence of large scale
fluctuations could alter the statistical
properties of the \lya\ forest or the bias of galaxies with respect
to the mass.  We computed conditional flux PDFs for galaxy samples
(selected on SFR) with a space density of $\Sigma=3\dunits$
in the two simulations.  We found only small differences in the results
(and therefore do not plot them), with a slight shift of the distribution
in the smaller box towards an increased number of low
decrement pixels and a corresponding dearth of high decrement pixels.
The conditional mean decrement is thus slightly depressed in the smaller
box, but again this impact is much smaller than the impact
of galaxy proximity itself.  We conclude that the lack of large scale
modes in the L11n64 simulation has only a minimal impact on the
conditional flux PDF for galaxies of this space density, and that
the box size of the L22n128 simulation should be adequate for our purposes.
The bias of rarer, more massive (or higher SFR) galaxies might be more
sensitive to the absence of long wavelength modes, but for rare
galaxies one needs a large simulation volume to get adequate statistics
in any case.

Figure~\ref{fig:resolution2} compares the results of the \bd and \bc
simulations, in the same format as Figure~\ref{fig:resolution}.  We
select galaxies above the same physical mass threshold, which
corresponds to $64\msph$ and $512\msph$, respectively.  The space
densities are $\Sigma \approx 3\dunits$.  In this comparison,
differences could arise from random differences in the structure
present, or from systematic effects of box size, mass resolution
(which is another factor of eight lower in L50n144), or our {\it
post-facto} imposition of a temperature-density relation in L50n144
(see \S\ref{sec:method}).  The good agreement in
Figure~\ref{fig:resolution2} shows that the combined impact of the
systematic effects must be reasonably small.  In what follows,
therefore, we will use L22n128 to make predictions for abundant galaxy
populations with $\Sigma > 3\dunits$ and L50n144 to make predictions
for rarer populations, where its larger volume is needed to provide
adequate statistics.  The differences in Figure~\ref{fig:resolution2}
are probably an upper limit on the level of systematic uncertainty
associated with the numerical parameters of the simulations. Even the
noticeable difference in the mean decrement for $\Delta\theta < 2'$ is
not highly significant relative to the statistical uncertainties,
since there are only 37 galaxies above the mass threshold in
the L22n128 simulation.  We conclude that the trends predicted by
our simulations are robust to the effects of finite resolution and
box size and that our quantitative results are only mildly sensitive
to these numerical effects.

\subsection{Predictions of the Conditional Flux PDF}
\label{sec:results}

Figure~\ref{fig:cpdf} presents the conditional flux PDF for the full
population of resolved galaxies ($M_b \geq 64\msph = 6.7\times 10^9
M_\odot$) in the L22n128 simulation.  There are 641 galaxies in the
box at $z=3$, and the space density of this population is $\Sigma =
48\dunits$.  (As indicated in the note to Table~\ref{tab:sims}, the
box is $18.74'$ on a side and has a redshift depth $\Delta z = 0.0379
= 2844\kms$.)  The upper left panel is for spectra smoothed over
$10\kms$, comparable to the spectral resolution of typical Keck HIRES
or VLT UVES spectra.  The unconditional flux PDF, shown by the solid
curve, reflects the probability distribution of gas overdensity
$P(\rho/\bar{\rho})$, which depends mainly on the amplitude of matter
fluctuations at $z=3$ and on the assumption that the primordial
fluctuations were Gaussian (see, e.g., the discussions by
\citealt{miralda96,miralda97,
cen97,rauch97,weinberg99a,croft99,mcdonald00}).  The shape of $P(D)$
is, of course, strongly influenced by the non-linear transformation
between density and flux decrement (eq.~\ref{eqn:fgpa}); for example,
the high and low density tails of $P(\rho/\bar{\rho})$ map to flux
decrements $D\approx 1$ and $D \approx 0$, respectively, so $P(D)$
peaks at these extremes.

Dot-dashed, dotted, and dashed curves show $P(D)$ for pixels that have a
galaxy at the same redshift within $\theta_{\rm max} = 2'$, $1'$, and $0.5'$,
respectively (comoving separations of 2.4, 1.2, and $0.6\hmpc$).
The preferential location of galaxies in overdense regions reveals
itself through shifts of the distributions towards higher flux
decrements, with corresponding increases in the mean decrement
(symbols along the top axis) and the fraction of saturated pixels
(symbols along the right axis), and decreases in the fraction of
transparent pixels (symbols along the left axis).  These shifts
become progressively stronger as the proximity to galaxies
increases ($\tmax$ decreases).  Pixels that contribute to
the $\Delta\theta < 0.5'$ histogram also contribute to the
$\Delta\theta < 2'$ histogram, but the number of galaxy--spectrum
separations is proportional to $(\Delta\theta)^2$, so the conditional
PDF for a given $\tmax$ is contributed largely by separations close to $\tmax$.
A sphere of radius $2.4\hmpc$ ($\Delta\theta=2'$) contains $3.4$
galaxies on average, so for the values of $\tmax$ in
Figure~\ref{fig:cpdf} the galaxy ``matched'' to any given pixel
is not necessarily the one with the smallest 3-d physical separation.
For the brighter, lower space density samples defined later,
the matched galaxy is usually also the closest one.

The remaining panels of Figure~\ref{fig:cpdf} show the unconditional
and conditional flux PDFs for spectra smoothed over 50, 100, and $200\kms$.
Smoothing brings the flux decrements in extreme pixels closer to the
mean, so for larger $R_s$ (which represents the $1\sigma$ width
of the Gaussian
smoothing filter) the fractions of saturated and transparent pixels
drop.  With $R_s=200\kms$ the unconditional PDF peaks close to
the mean decrement, though it remains fairly broad.  Roughly speaking,
the flux PDF of smoothed spectra reflects the PDF of density smoothed
on a similar physical scale, but the smoothing of the spectra is
1-d rather than 3-d, and smoothing does not commute with the
non-linear transformation between density and flux.

For our purposes, the important feature of these panels is that the
difference between the unconditional and conditional flux PDFs, and
between the conditional PDFs for different $\tmax$, persists for
smoothing lengths up to $200\kms$, and is perhaps even enhanced
relative to the minimal smoothing case.  Since we see a strong
signature of galaxy proximity at $\Delta\theta=2' = 304 \kms$, it is
no surprise that the signature persists when spectra are smoothed over
this physical scale; the difference between the unconditional and
conditional PDFs reflects the large scale environments in which
galaxies reside.  Figure~\ref{fig:cpdf} shows that the simulation
predictions can be tested with moderate-to-low resolution spectra,
thus allowing the use of relatively faint AGN as targets for
observational studies. 

The principal difficulty with lower resolution spectra is the greater
difficulty of continuum fitting, which causes systematic underestimation
of flux decrements, especially at higher redshifts where large expanses
of unabsorbed continuum are rare.  However, the qualitative trends
with galaxy proximity should persist even in the presence of continuum
fitting errors.
With high resolution
spectra, the dependence of the conditional flux PDF (or properties
like $\meand$, $\fsat$, and $\ftrans$) on smoothing scale can be used
to test theoretical predictions, in addition to the dependence on
galaxy proximity itself.  Note that the mean of the unconditional flux
PDF does not depend on $R_s$, since smoothing preserves flux overall,
but the mean decrement for conditional PDFs decreases with increasing
$R_s$, since pixels close to galaxies tend to have high decrements
that are reduced by smoothing.

Another indicator of the physical scale of correlations is the dependence
of the conditional PDF on the velocity separation between the pixel
redshift and the galaxy redshift.  In Figure~\ref{fig:deltav},
long-dashed lines and solid symbols repeat results for the case
investigated in Figures~\ref{fig:resolution}--\ref{fig:cpdf},
velocity separation $\Delta V=0$.  Upper and lower panels show
$\tmax=2'$ and $0.5'$, and left and right columns show smoothing
lengths $R_s=10$ and $50\kms$.  Dashed, dotted, and dot-dashed lines
show the conditional PDFs computed from pixels that are offset
by $|\Delta V| = 100,$ 200, and $400\kms$ from the galaxy redshift, where we have averaged the contributions at $\pm \Delta V$.
For $|\Delta V| = 100$ or $200\kms$, the signature of galaxy proximity
on the flux PDF persists, but it is significantly weakened.
By $|\Delta V| = 400\kms$, it has mostly vanished.  The dependence
on velocity separation provides another potential test of the simulation
predictions.  Perhaps more significantly, Figure~\ref{fig:deltav}
shows that redshift measurement errors of $\sim 100-400\kms$ would
significantly affect the shape of the conditional flux PDF.
We will return to this point in \S\ref{sec:zerror} below.

The unconditional flux PDF is weighted by volume, since every volume element
has an equal chance of becoming a pixel in a \lya\ forest spectrum.
Therefore, even if galaxies traced the underlying mass distribution,
we would expect the conditional flux PDF to differ from the unconditional
PDF because the mass itself is preferentially located in overdense regions.
Figure~\ref{fig:bias} demonstrates, however, that the strong dependence
of the flux PDF on galaxy proximity is primarily a result of
galaxy bias, not the difference between volume and mass weighting.
Light curves, repeated from the corresponding panels of Figure~\ref{fig:cpdf},
show the unconditional PDFs and the conditional PDFs for galaxy separations
$\Delta \theta < 2'$ and $\Delta \theta < 0.5'$.  Heavy curves show
conditional PDFs in which we use
randomly selected dark matter particles in place
of galaxies.  While these curves are shifted relative to the unconditional
PDFs shown by the solid lines, they are not shifted nearly as much as the
galaxy conditional PDFs.  High-redshift galaxies form in special locations
of the large scale matter distribution, and the signature of these
special environments is imprinted on the nearby \lya\ forest.

\subsection{Dependence on Galaxy Properties}
\label{sec:properties}

In the local universe, galaxies with different stellar populations,
star formation rates, morphologies, and masses tend to populate
different environments, and the same may be true of high redshift
galaxies.  Theoretical models predict some dependence, albeit
rather weak, of the LBG autocorrelation function on galaxy mass
or SFR \citep[e.g.,][]{mof96,katz99},
and observations provide at least tentative evidence for stronger
clustering of brighter LBGs \citep[e.g.,][]{adelberger98,giavalisco01}.
The dependence of the conditional flux PDF on the properties of the
conditioning galaxies is a potential diagnostic for the dependence
of LBG properties on large scale environment.  Here we will
focus on instantaneous SFR for defining different galaxy subsamples,
since it is the property most closely tied to rest-frame UV luminosity.
In the simulated galaxy population, these SFRs are correlated with
galaxy baryonic mass \citep{weinberg02b}, so this procedure is
approximately, but only approximately, equivalent to selecting on
baryonic mass.

In Figure~\ref{fig:luminosity}, heavy solid lines show the conditional
flux PDFs for the full galaxy population of the L22n128 simulation,
as in Figure~\ref{fig:cpdf}, for $\Delta\theta < 2'$ (top)
and $\Delta \theta < 0.5'$ (bottom), with spectral smoothing
$R_s=10\kms$ (left) or $50\kms$ (right).  Light solid lines
show the corresponding results for the full population of resolved
galaxies ($M_b > 64\msph = 5.4\times 10^{10}M_\odot$)
in L50n144.  The space density of this more massive galaxy
population is $\Sigma=3\dunits$, with 450 galaxies in the
simulation volume.  Dotted lines show results for the 150
galaxies with the highest SFRs, and dashed lines show results
for the top 50 galaxies.  The conditional flux PDF is remarkably
insensitive to galaxy SFR over a range of nearly 150 in space
density, though for $\Delta\theta < 0.5'$
the more massive, high SFR galaxies do show
slightly higher mean decrement and saturated fraction and slightly
lower transparent fraction, as expected
if they reside in denser environments.

One of the simplest and most widely used models for the spatial
clustering of LBGs associates each LBG with a single dark matter
halo \cite[e.g.,][]{mof96,adelberger98,moscardini98,bagla98}.
The large scale bias factor of these halos can be computed using
the approximations of \cite{mow96} and \cite{sheth01}.
\cite{katz99} show that the correlation function of $z=3$ galaxies
in SPH simulations like these is similar to that of dark matter
halos of the same space density, even though the galaxy-halo
association for a given space density is not truly 1-to-1.
Figure~\ref{fig:halo} presents an analogous comparison for the conditional
flux PDF.  Light lines and symbols show results for the 450 resolved
galaxies in the L50n144 simulation, while heavy lines and symbols
show the results when these galaxies are replaced by the 450 most
massive dark matter halos in the simulation (identified by a
friends-of-friends algorithm with linking parameter of 0.2).
The conditional PDFs for galaxies and halos are very similar,
with galaxies showing a slightly stronger preference for regions
of high flux decrement, suggesting that they are more likely to
occur in halos with higher density environments (where the halos
themselves tend to be more massive).  This similarity holds for
other choices of angular separation, spectral smoothing scale,
and population space density, not shown in Figure~\ref{fig:halo}.

A more general model for galaxy bias is the ``halo occupation distribution''
(HOD) formalism, which characterizes the mean number of galaxies of a
given type in halos of virial mass $M$ by an arbitrary function
$N_{\rm avg}(M)$.  This approach has been widely applied to studies
of low redshift clustering \citep[e.g.,][]{jing98,seljak00,ma00,peacock00,
scoccimarro01,berlind02,cooray02a,cooray02b,marinoni02,scranton02,zehavi03},
and \cite{bullock01} and
\cite{moustakas02} have analyzed LBG clustering in this context.
If we neglect the finite size of the halos themselves, then the
influence of different $N_{\rm avg}(M)$ choices on the conditional
flux PDF can be calculated by simply weighting the contribution of
each halo by $N_{\rm avg}(M)$.  Figure~\ref{fig:hod} shows the
conditional flux PDF based on randomly selected dark matter particles
(solid line, as in Figure~\ref{fig:bias}) and on the top 450 halos of the
L50n144 simulation, with contributions weighted by $W\propto M^0$ (one
galaxy per halo), $W\propto M$ (galaxy number proportional to halo mass),
and additional cases $W\propto M^{0.5}$ and $W\propto M^{1.5}$.
While the bias of halos with respect to dark matter remains clearly
evident, the conditional flux PDF is almost completely independent
of the relative weighting of the halos, and, by implication, of the
galaxy HOD over the rather wide parameter range considered here.
Similar results hold for other angular separations and spectral smoothings.
Presumably the more massive halos do tend to reside in denser
environments, but saturation removes the sensitivity of the flux
decrement to the neutral hydrogen density once it becomes sufficiently high.
Thus, the sign of the effect, more absorption when
more massive halos are weighted more strongly, is in the expected direction,
but the magnitude is tiny.

At one level, the results in Figures~\ref{fig:luminosity}--\ref{fig:hod}
are disappointing, since they suggest that the conditional flux PDF
will not help to distinguish among competing scenarios for the origin
of LBG properties.  However, the robustness of the predictions
to the details of the galaxy formation assumptions makes the conditional
flux PDF a good diagnostic for the influence of galaxies on the IGM.
Since a wide range of reasonable galaxy bias models make similar predictions,
a substantial departure from these predictions would most likely arise
from galaxies directly affecting the \lya\ forest through photoionization,
winds, or other processes.  Furthermore, differences in the conditional
flux PDF for different galaxy classes --- based on luminosity, stellar
populations, or emission line profiles, for example --- would probably
reflect their differing effects on the IGM, not their different
clustering properties.

\subsection{The Conditional Mean Flux Decrement}
\label{sec:summary}

The dependence of the mean flux decrement on the galaxy proximity
condition provides a convenient summary of our results.
The mean decrement does not capture the full information in
the PDF; we have tried, for example, scaling the optical
depths of the unconditional flux PDF by a constant factor to see whether we
can recover the conditional flux PDF simply by matching its mean,
and we cannot.  However, measurement of the full conditional flux PDF
requires many galaxy-quasar pairs, so measurements of the conditional
mean decrement are likely to come much sooner (e.g., ASSP).

Figure~\ref{fig:mean} plots the mean flux decrement as a function of
maximum angular separation $\tmax$ for galaxy populations of different
space density, summarizing and reinforcing the results in
Figures~\ref{fig:cpdf} and~\ref{fig:luminosity}.
The spectral smoothing length is $R_s=10\kms$.
Hexagons and pentagons show results for the full populations
of resolved galaxies in the L22n128 and L50n144 simulations,
with space densities $\Sigma=48\dunits$ and $3\dunits$, respectively.
Squares and triangles show results for the 150 and 50 ``brightest''
(i.e., highest SFR) galaxies in L50n144.  For $\Delta\theta < \tmax = 4'$
($4.8\hmpc$ comoving)
the mean decrement is already well above the global mean of 0.36, and
it rises steadily with decreasing $\tmax$ as the \lya\ spectra
probe the denser IGM in the immediate vicinity of the galaxies,
with $\meand > 0.9$ for $\tmax=0.25'$.  The mean decrement varies
only weakly with SFR, with the rare, high SFR galaxies having slightly
higher mean decrement.
The $\times$'s show, for the full population of L50n144, the conditional
mean decrement computed in bins of angular separation
$0.8\tmax < \Delta\theta < 1.25\tmax$, instead of all separations
$\Delta\theta < \tmax$.  This ``differential'' form of the statistic
shows a similar trend, but the increase of $\meand$ at a given $\tmax$
is, of course, smaller, since closer separation pairs have been eliminated.

Figure~\ref{fig:mean2} plots the mean decrement against $\tmax$ for
random dark matter particles and for halos weighted in various
ways, as in Figures~\ref{fig:bias} and~\ref{fig:hod}.
The effect of bias stands out clearly as an offset between
the matter and halo results.  Results for halos are similar to
those for galaxies in Figure~\ref{fig:mean}, though the mean
decrements near halos are typically slightly lower.  The weighting
of halos has only a small impact on the conditional mean
decrement, with a modest increase in $\meand$ for a stronger
weighting of high mass halos.

\subsection{Local Photoionization}
\label{sec:gpi}

We have so far assumed that the
photoionizing background is uniform, and \cite{croft02b} show that
this assumption should be adequate for most computations of \lya\
forest statistics.  However, the conditional flux PDF is derived from
the small fraction of pixels that lie close to galaxies, and if
the galaxies make a significant contribution to the ionizing background,
then the flux decrement in these pixels may be depressed by the
ionizing flux of their nearest galaxy neighbors.
With the ionizing background predicted by \cite{haardt96} based on
the observed quasar population, these simulations (and similar ones
by other groups) already match the observed mean decrement of the
\lya\ forest given a baryon density $\Omega_b=0.02h^{-2}$.  However,
estimates of the background based on the proximity effect usually
yield a higher intensity than predicted from the quasar population
alone \citep[][and references therein]{scott00}, and observations
of LBGs suggest that the escape fraction of ionizing photons is
high enough to make them an important contributor to the UV background
\citep{steidel01}.  Within the theoretical and observational uncertainties,
there is probably room for galaxies to make a contribution equal to that of
quasars without making the predicted mean flux decrement too low at
$z\sim 3$ (see \citealt{schirber03} for a recent discussion of these issues).

The intensity of the radiation from an individual galaxy or quasar
falls off as $1/r^2$, but it is further attenuated by redshifting
and, more importantly for the case of ionizing radiation, by
IGM absorption.  Haardt \& Madau (\citeyear{haardt96}; see also
\citealt{madau99}) estimate that a path length $\Delta z \sim 0.17$
produces an optical depth of $\tau(912\hbox{\AA})\sim 1$ at $z=3$,
corresponding to $100\hmpc$ (comoving) for our cosmology.
\cite{fardal93} conclude that the effective attenuation length
for ``average'' ionizing photons is $\sim 2.4$ times that at 912\AA,
so a reasonable estimate (with probably a factor of two uncertainty)
is $r_{\rm att} \sim 240\hmpc$ (comoving).
Here we will make the simple approximation that intensity falls
as $1/r^2$ until $\rmax=r_{\rm att}$ and is sharply truncated
beyond $\rmax$.
If a population of galaxies with comoving space density $n$
and mean ionizing luminosity $\meanl$ contributes a fraction $f$
of the ionizing background intensity $I$, we then have
\begin{equation}
fI ~=~ \int_0^{\rmax} 4\pi r^2 dr \,n \,{\meanl \over 4\pi r^2}
   ~=~ n\meanl \rmax ~,
\label{eqn:fI}
\end{equation}
implying
\begin{equation}
I ~=~ {n\meanl\rmax \over f} ~.
\label{eqn:I}
\end{equation}
The intensity from a galaxy with luminosity $L_i$ is equal to the mean
background intensity at an ``influence radius'' $r_i$ given by
\begin{equation}
{L_i \over 4\pi r_i^2} = I \qquad \Longrightarrow \qquad
r_i = \rmax \times \left[ {L_i \over \meanl}
      {f\over 4\pi r^3_{\rm max} n}\right]^{1/2} ~.
\label{eqn:ri}
\end{equation}
Note that $r_i$ decreases as either $\rmax$ or $n$ increases,
since either change makes
the contribution of a nearby galaxy less important relative to that
of the numerous, distant galaxies.

For $f=0.5$, $L_i=\meanl$, $\rmax=240\hmpc$, and
$n=1.2\times 10^{-3} h^3 {\rm Mpc}^{-3}$
($\Sigma = 1\dunits$), equation~(\ref{eqn:ri}) yields $r_i=0.37\hmpc$
(all length scales comoving).
For our cosmological model, the corresponding angular and
velocity scales are $\Delta\theta = 0.3'$, $\Delta V = 47\kms$, respectively.
This simple estimate suggests that galaxy photoionization could
have a significant impact on the conditional flux PDF at separations of
$\Delta\theta \sim 0.5'$, though its impact at $\Delta\theta \sim 2'$
is likely to be small.
The optical depth of gas at temperature $T$, with electron density $n_e$,
and at a distance $r$ from a galaxy is reduced relative to the optical depth
$\tau_u$ of the uniform background case by a factor
\begin{equation}
\tau/\tau_u =
  \left[1+\fgamma\left({r_i\over r}\right)^2\right]^{-1} ~,
\label{eqn:tau}
\end{equation}
where
\begin{equation}
{1 \over \fgamma} = {\Gamma_u + \Gamma_c(T) n_e \over \Gamma_u} \approx
  1 + 0.29(\rho/\bar{\rho})e^{-1.578/T_5} T_5^{1/2} (1+T_5^{1/2})^{-1}
\label{eqn:fgamma}
\end{equation}
is the ratio of the total (photo $+$ collisional) ionization rate
to the uniform photoionization rate $\Gamma_u$.
The second part of equation~(\ref{eqn:fgamma}) incorporates the
\cite{haardt96} value of $\Gamma_u=8.3\times 10^{-13}\, {\rm s}^{-1}$
and the \cite{cen92} expression for collisional ionization
at temperature $T=10^5 T_5\K$ (which is also used in TreeSPH
and TIPSY), assuming fully ionized gas to relate
$n_e$ to $\rho/\bar{\rho}$.

Figure~\ref{fig:gpi} illustrates the impact of galaxy photoionization
on the conditional mean decrement (left panels) and saturated fraction
(right panels) for the $\Sigma=1\dunits$ sample from the L50n144 simulation.
We compute the influence radii via equation~(\ref{eqn:ri}) assuming that each
galaxy's ionizing flux is proportional to its SFR and that
galaxies of this space density collectively produce a fraction $f=0.5$ of
the ionizing background, with fainter galaxies having no significant
ionizing flux.
Asterisks show the result of a simplified calculation in which we
multiply the optical depths of the extracted spectra by the
factor in equation~(\ref{eqn:tau}) assuming $\fgamma=1$ and a pixel-galaxy
distance $r$ corresponding to the transverse
separation between the galaxy and the line of sight.
As expected from the order-of-magnitude estimate above, the
suppression relative to the uniform UV background case
is small at $\Delta\theta \geq 1'$, but
the mean decrement and saturated pixel fraction are noticeably
reduced for $\Delta\theta \leq 0.5'$.
Triangles show the results of a complete
calculation, in which we put individual galaxy UV sources into
the simulation before extracting the \lya\ forest spectra with TIPSY,
applying equation~(\ref{eqn:tau}) to each SPH particle with $r$
equal to
the galaxy-particle separation and $\fgamma$ computed
from the particle's temperature and density.
Here the reduction of $\meand$ and $\fsat$ is much smaller, and the
predicted trends no longer turn over at the smallest angular separations.

Why does the approximate calculation drastically overestimate the impact
of galaxy photoionization?  One possibility is that collisional ionization
in the hotter, denser gas within $r\sim 0.5\hmpc$ of galaxies
dilutes the effect of photoionization by making $\fgamma \ll 1$.
However, if we repeat the TIPSY calculation with $\fgamma=1$ for all
particles, we get a nearly identical result for the conditional
mean decrement.  Collisional ionization {\it is} important along some lines
of sight (compare equation~[\ref{eqn:fgamma}] to Figure~\ref{fig:profiles}),
but these are cases where dense gas produces heavily saturated absorption,
which remains saturated
even if the optical depth is reduced by a factor of several.
Thus, overestimating photoionization by setting $\fgamma=1$ still
leaves a flux decrement $D\approx 1$.
The second possibility is that the gas producing absorption at the
galaxy redshift is further away than the transverse separation $r_t$,
and is shifted to the galaxy redshift by its peculiar infall velocity.
Figure~\ref{fig:dlos} shows the mean line-of-sight distance,
weighted by optical depth, of gas that produces absorption within
$20\kms$ of the galaxy redshift.  We consider only galaxy-spectrum
pairs with $\Delta\theta<0.5'$, implying a transverse separation
$r_t<0.6\hmpc$ (comoving); results are similar for $\Delta\theta<0.25'$.
Figure~\ref{fig:dlos} shows that absorption for these close pairs
arises mainly in
gas whose line-of-sight distance substantially exceeds the influence
radius $r_i$ and is therefore little affected by galaxy photoionization.
When the optical depth is high ($\tau>10$, left-hand panel),
the mean line-of-sight distance is usually smaller, but in these cases
even a factor of several reduction in $\tau$ does
not move the flux decrement significantly below $D=1$.
Thus, even if galaxies contribute a large fraction of the ionizing
background, local photoionization has little effect
on the conditional mean decrement or saturated fraction
because the real-space distance to absorbing gas is larger
than the redshift-space distance, except for lines of sight that
are heavily saturated.

\cite{steidel02} find that $\sim 3\%$ of the galaxies in their
LBG sample exhibit detectable AGN activity.
Given the apparent ubiquity of supermassive black holes in local
galaxies, it seems plausible that most LBGs go through AGN phases,
with $\fq \sim 0.03$ being the ``duty cycle'', i.e. the fraction
of time that these black holes are active at $z\sim 3$.
Could photoionization by these low luminosity AGN have a larger
impact on the local IGM than the galaxies themselves?
If the interval between periods
of AGN activity is longer than the time
\begin{equation}
\teq = {n^{\rm eq}_{\rm HI} \over \alpha(T) n_e n_{\rm HII}} =
  [\Gamma_u + \Gamma_c(T) n_e]^{-1} \approx 4\times 10^4 \fgamma\,{\rm yrs}
\label{eqn:teq}
\end{equation}
that it takes gas to return to its equilibrium neutral hydrogen fraction
after being fully ionized,\footnote{We assume that the equilibrium
state is already highly ionized, so that the addition of
$n^{\rm eq}_{\rm HI}$ hydrogen ions
does not significantly increase $n_{\rm HII}$
or $n_e$.  The second equality follows from the equilibrium condition
$\alpha(T) n_e n_{\rm HII} = \Gamma_u n^{\rm eq}_{\rm HI}
+ \Gamma_c(T) n_e n^{\rm eq}_{\rm HI}.$}
then we are essentially back to the uniform ionization case:
the few LBGs that currently host AGN have large ``proximity zones,''
but the rest have their normal complement of associated \lya\ optical
depth.  However, if the process of feeding gas to the central
black hole is sufficiently stochastic, then AGN might ``flicker''
on and off on timescales shorter than $\teq$, with individual
activity cycles lasting $\tactive \la \fq \teq \la 10^3\,$years.
In this case, gas around a large fraction of LBGs could be
out of ionization equilibrium, with the neutral hydrogen fraction
depressed by the memory of the most recent AGN outburst.

Suppose we assume that the AGN associated with the galaxies under
study collectively produce 50\% of the ionizing background.
Since the time-averaged flux of this population equals, by
assumption, the flux that we previously ascribed to the galaxies
themselves, the photoionization rate at distance $r$ during an
active phase must exceed the previous value by a factor $\fq^{-1}$,
i.e., $\Gammaq = \Gamma_u(r_i/r)^2\fq^{-1}$.
During an active phase, the neutral hydrogen density at distance $r$
drops exponentially in time,
$n_{\rm HI} \approx n^{\rm eq}_{\rm HI} \exp(-\Gammaq t)$,
until it reaches a new equilibrium value with \lya\ optical depth
\begin{equation}
\tau/\tau_u =
  \left[1+\fq^{-1}\fgamma\left({r_i\over r}\right)^2\right]^{-1} ~.
\label{eqn:tau2}
\end{equation}
However, at most radii the duration of the active phase will
not be long enough to achieve this equilibrium.  If we assume that
the activity during a recombination interval $\teq$
occurs in a single ``outburst'' of duration $\tactive=\fq\teq$,
then the optical depth at the end of this phase is
\begin{equation}
\tau/\tau_u \approx \exp(-\Gammaq \tactive) =
  \exp\left[-\fgamma\left({r_i \over r}\right)^2\right]~,
\label{eqn:tau3}
\end{equation}
except that it never falls below
the equilibrium value of equation~(\ref{eqn:tau2}).
Note that the duty cycle $\fq$ cancels out of equation~(\ref{eqn:tau3})
because we have fixed the time-averaged flux of the AGN population
relative to $\Gamma_u$.  If the AGN are more luminous while they
are on, then they must be active for less time, producing the same number
of ionizing photons.  We see from equation~(\ref{eqn:tau3})
that the scale over which recurrent AGN activity can affect
the conditional mean flux decrement is the same influence radius
that we found previously, but that departures from photoionization
equilibrium allow the impact within this radius to be much stronger.

Since we have already found that much of the absorption for small
$\Delta\theta$ comes from gas beyond the influence radius, we can guess
from this analysis that local AGN photoionization will have little
impact on the conditional mean decrement.
Filled circles in Figure~\ref{fig:gpi} confirm this expectation, showing
a calculation based on
equation~(\ref{eqn:tau3}), with sources embedded in the
TIPSY spectral extraction and the same influence radii used
previously.  The minimum optical depth is given by equation~(\ref{eqn:tau2})
with $\fq^{-1}=30$.  While the absorption is slightly suppressed
relative to the galaxy photoionization case, the mean decrement and
saturated fraction continue to increase as $\tmax$ decreases.
These results can be considered a conservative
upper limit on the effects of recurrent AGN activity, since we
have ignored the partial return to equilibrium
with the uniform background that will occur following any given active cycle.
If we attributed 100\% of the ionizing background to the AGN
associated with the observed LBGs, then the influence radii would be
larger by $2^{1/2}$, but the impact on the conditional mean decrement
would be only slightly stronger.

We conclude that local photoionization by galaxies or by the
AGN that they host will not reverse the trend of increasing \lya\ forest
absorption with decreasing angular separation.
Figure~\ref{fig:dlos} suggests that {\it any} feedback mechanism must
have a strong influence out to comoving distances $\sim 1\hmpc$
or more if it is to effect such a reversal.
We have examined simple models in which we completely eliminate
neutral hydrogen out to a fixed radius around target galaxies or, 
alternatively,
reduce the neutral fraction within this radius by a factor of three.
If we eliminate all neutral hydrogen to $r=1\hmpc$, then the conditional
mean decrement drops to $\meand \approx 0.6$ for $\tmax=0.25'$ and $0.5'$,
and the saturated pixel fraction falls to $\fsat \approx 0.4$.
However, a factor of three reduction out to $1\hmpc$, or complete
elimination out to $0.5\hmpc$, has only small impact, comparable to
that of the AGN ionization model shown in Figure~\ref{fig:gpi}.
Elimination of neutral hydrogen out to $0.75\hmpc$ has an intermediate
effect, reducing $\meand$ to $\sim 0.75$ and $\fsat$ to $\sim 0.5$.

\subsection{Redshift Errors}
\label{sec:zerror}

We showed in Figure~\ref{fig:deltav} that the signature of the
overdense environments of galaxies in the conditional flux PDF is
substantially weakened if one looks at pixels $\sim 200\kms$
away from the galaxy redshift instead of at the galaxy redshift
itself.  (The angular separation corresponding to $Hr=200\kms$
is $1.3'$.)  Since the redshifts of LBGs are usually estimated
from nebular emission or absorption lines, which could be offset
relative to the mean systemic velocities, this result suggests that
redshift measurement errors could have a significant impact on
practical determinations of the conditional flux PDF.
To address this point quantitatively, we show in Figure~\ref{fig:zerror}
the conditional mean decrement and saturated fraction when pixel
redshifts are drawn from a Gaussian of mean zero and dispersion
$\sigma_{\Delta V}=150\kms$ (squares) or $300\kms$ (triangles).
We use the $\Sigma=1\dunits$ sample of the L50n144 simulation,
with no galaxy photoionization.  Redshift errors of
$\sigma_{\Delta V}=150\kms$, similar to those estimated by ASSP,
depress $\meand$ and $\fsat$ by $\sim 0.05$ at $\tmax=2'$ and by
$\sim 0.1-0.2$ at $\tmax \leq 1'$.  Errors of $\sigma_{\Delta V}=300\kms$
depress $\meand$ and $\fsat$ more severely, and they largely (though
not entirely) remove the trend of increasing absorption with
decreasing $\tmax$.
Redshift errors also reduce the already small
differences between models with and without galaxy photoionization,
since the absorption is now usually measured at a pixel further
away from the galaxy.

\section{Large Scale Correlations}
\label{sec:cube}

Inspired by the observational work of ASSP and ASPS, we also
examine a statistical measure that characterizes LBG-\lya\ forest
correlations on larger scales than those examined in \S\ref{sec:cpdf}.
We divide each of the 1200 random spectra through the L50n144
simulation into $L/S$ segments of comoving length $S$, where
$L=50\hmpc$ is the box size.
For each segment $i$, we compute the mean flux decrement
$D_{S,i}$ along the segment, and we count the number of
galaxies $N_i$ in a cube of side length $S$ centered on the
spectral segment, making use of the simulation's periodic boundaries
as necessary.
We include galaxy peculiar velocities when computing positions
along the line of sight, again utilizing the periodic boundaries
when needed.
We then consider all segments in a bin of $\meands$
values and compute the mean value of the galaxy density contrast
$\delta_{{\rm gal},i} = N_i/\meann -1$,
where $\meann=S^3n$ is the average number of galaxies in
a cubic cell of size $S$.  We use the full resolved galaxy population
of the simulation, for which $\meann = 3.6(S/10\hmpc)^3$.
Poisson noise fluctuations in $\dgal$ are large,
and if one plots the mean value of $\meands$ in bins of $\dgal$,
then a ``Malmquist''
type bias effectively erases any trend (there are more cells of
intrinsically low density contrast that scatter to high $\dgal$
than vice versa).
However, $\meands$ is not a noisy quantity, and if one has many
cells, then Poisson fluctuations in $\dgal$ average to zero without bias.
A plot of $\dgal$ in bins of $\meands$ therefore reveals the
preferential location of galaxies in overdense environments.

The solid points in the
upper panels of Figure~\ref{fig:cube} show $\dgal$ vs.\ $\meands$
for $S=10\hmpc$ (left) and $5\hmpc$ (right).  Error bars represent
the $1\sigma$ error on the mean, i.e., the dispersion of
the $N_{\rm seg}$ values of $\dgal$ in the $\meands$ bin
divided by $(N_{\rm seg}-1)^{1/2}$.  Smooth curves show
the cumulative distribution of $\meands$ values.  Only a small number
of segments have $\meands>0.6$ for $S=10\hmpc$,
so the error bars on $\dgal$ become large for high $\meands$
values.  Nonetheless, there is a clear, strong trend of increasing
$\dgal$ with increasing $\meands$ up to $\meands \approx 0.65$,
where the mean galaxy overdensity is $1+\dgal \approx 2.5$.
For $S=5\hmpc$, there are more high decrement segments and the
trend of $\dgal$ with $\meands$ continues to $\meands \approx 1$.

The open points in Figure~\ref{fig:cube} show the results using dark matter
particles instead of galaxies.  While there is a steady trend of
$\ddark$ with $\meands$, the bias of the galaxy population is
strikingly evident in the much steeper slope of the $\dgal-\meands$
trend.  The suppression of high redshift galaxy formation in
underdense regions is especially clear,
with mean density contrasts $\dgal$ approaching $-1$
for $\meands<0.2$.  In the $S=5\hmpc$ plot, the physical requirement
that $\delta \geq -1$ forces the galaxy correlation to be concave
upward at low $\meands$.
Peculiar velocities make the $\ddark-\meands$ correlation significantly
steeper in redshift space than in real space, as coherent flows
amplify density contrasts \citep{sargent77,kaiser87}.
However, the $\dgal-\meands$ correlation is minimally affected by
peculiar velocities owing to the high bias (and correspondingly
low $\Omega_m^{0.6}/b$) of the galaxy population.

The rms fluctuations of $\meands$ and density contrast are
$(\sigD,\sigdm,\siggal)=(0.13,0.35,1.01)$
for $S=10\hmpc$ and $(0.17,0.56,1.93)$ for $S=5\hmpc$.
Here we have subtracted a shot noise contribution to the galaxy
dispersion assuming Poisson statistics, i.e., we define
$\sigma_{\rm gal}=
\left[\langle (N-\meann)^2 \rangle -\meann \right]^{1/2}/\meann$,
where $\langle ...\rangle$ represents an average over all cells
and the division by $\meann$ converts from a dispersion of $N$ to
a dispersion of $\dgal$.  We also subtract shot noise when
computing $\sigdm$, but since the number of dark matter particles
is large, the correction is negligible.

The ratio of correlation slopes between $\dgal-\meands$ and
$\ddark-\meands$ is roughly the ratio of $\siggal$ to $\sigdm$,
as we show in the bottom panels of Figure~\ref{fig:cube}, which
plot $\dgal/\siggal$ or $\ddark/\sigdm$ against $(\meands-\bar{D})/\sigD$,
where $\bar{D}=0.36$ is the unconditional mean decrement.
In these normalized variables, the galaxy and dark matter correlations
are similar, though the constraint $\delta \geq -1$ forces the
galaxy correlation to become shallower at low $\meands$, while
the dark matter trend remains linear down to essentially zero
flux decrement.  Thus, linear bias appears to be a good but not
perfect approximation for this statistic.  Our results for dark matter
agree well with those of \cite{mcdonald02}, who examine the unnormalized
and normalized correlations of $\ddark$ with $\meands$ using the
hydro-pm approximation of \cite{gnedin98}.
For $10\hmpc$ cubes, they find a correlation with a slope of 0.6 in
normalized variables, bending towards shallower slopes for
$(D_{10}-\bar{D})/\sigD \ga 1$ ($\tilde{\delta}_F \ga 1$ in their
notation).  The box diagonals of the lower panels of Figure~\ref{fig:cube}
have a slope of 0.6, so one can see by visual inspection that we
obtain a similar slope and turnover, and that the
slope for $5\hmpc$ is nearly the same as for $10\hmpc$.
\cite{mcdonald02} show that variations in the \lya\
forest assumptions --- e.g., the temperature-density relation, the
mean decrement, or the flux power spectrum --- have only a small effect
on the $\ddark-\meands$ correlation when it is plotted in
normalized variables, so we expect the same would hold true for the
$\dgal-\meands$ correlations shown here.

As a quantitative measure of correlation strength, we compute the
cross-correlation coefficients
$\rgalD \equiv \langle \dgal \delta_D \rangle / (\siggal\sigD)$ and
$\rdarkD \equiv \langle \ddark \delta_D \rangle / (\sigdark\sigD)$,
where $\delta_D \equiv \meands-\bar{D}$, and the averages are
over all cells/segments.  At $10\hmpc$, the correlation coefficients
are 0.49 and 0.66 for galaxies and dark matter, respectively.
The coefficients for dark matter are substantially
higher in redshift space than they would be in real space, while
for galaxies they are nearly identical in redshift space and
real space.  The cross-correlation between galaxies and dark matter
is high, $\rgaldm\approx 0.95$, and would again be lower in real space.
However, $\rgalD$ is lower than $\rdarkD$ for both cube sizes,
perhaps because the curvature of the $\dgal-\ddark$ relation at
low $\delta_D$ suppresses the contribution of these cells to the
correlation coefficient.  As emphasized by ASPS, the
correlation $\rdarkD$ is necessarily below unity because the
flux decrement $\meands$ gives a (non-linearly) weighted average
of the density along the line of sight through the cube center,
and this quantity is imperfectly correlated with the mean density
of the cube as a whole.

Figure~\ref{fig:cubehod} resembles Figure~\ref{fig:cube}, but here the
$\delta-\meands$ correlation for galaxies is compared to that for the
450 most massive friends-of-friends halos, with the same weighting
schemes used in Figures~\ref{fig:halo} and~\ref{fig:hod}.
The correlation is insensitive to halo weighting except at high $\meands$,
since massive halos are rare in lower density regions and assigning
them higher weights makes little difference.
At high $\meands$, the contrast is usually stronger when massive
halos receive more weight, as one would expect, but our results in
this regime are fairly noisy.  The rms density contrast $\sigma$ {\it is}
sensitive to the relative halo weighting, so in normalized variables
(bottom panels) the correlation is actually shallower when massive
halos are more strongly weighted, and correlation coefficients are
correspondingly lower.  The galaxy correlations most closely resemble
those for equally weighted halos or halos with weight $\propto M^{0.5}$,
and both the rms fluctuations of galaxies and their correlations with
\lya\ flux decrement are lower than they would be if the number
of galaxies were proportional to halo mass.

\section{Discussion}
\label{sec:discussion}

The conditional flux PDF considered in \S\ref{sec:cpdf} and
the $\dgal-\meands$ correlation considered in \S\ref{sec:cube}
offer empirical tools with which to test the prediction that
high redshift galaxy formation is strongly biased, and to search for
the influence of star-forming galaxies on the surrounding IGM.
The predicted flux PDF shifts systematically towards higher absorption
in pixels close to galaxies --- the smaller the angular separation,
the higher the mean decrement and saturated fraction and the
lower the transparent fraction.
The shift of the flux PDF remains strong in spectra
smoothed over 50, 100, or $200\kms$, allowing further tests of
the predicted scale of correlations and demonstrating that moderate
resolution spectra of relatively faint AGN
may profitably be used in such investigations.
The influence of galaxy proximity is much stronger than that of
random mass elements (dark matter particles), a clear signature
of the preferential formation of galaxies in overdense environments.

The conditional flux PDF is only weakly sensitive to the
baryonic mass or SFR of the sample galaxies over the range that
we have investigated, spanning more than two orders of magnitude
in space density.  It is also insensitive to the occupation
distribution of dark matter halos.  While this insensitivity
reduces the utility of this statistic as a diagnostic of galaxy
formation, it increases its power as a diagnostic of galaxy
feedback, since the no-feedback prediction is robust.
We have investigated the potential impact of one feedback process,
galaxy photoionization, and find that it has only a small
impact on the conditional mean flux decrement, even if the bright
galaxies used in the conditional PDF measurement
contribute a large fraction of
the ionizing background.  The local flux from these galaxies
is comparable to the uniform background at ``influence radii''
$r_i \sim 0.4\hmpc$ (comoving), corresponding
to angular separations $\Delta\theta \sim 0.3'$,
and in the absence of peculiar velocities
this flux would drive down the average absorption at smaller separations.
However, the gas producing absorption at the galaxy redshift
is typically $\ga 1\hmpc$ away along the line of sight, where
it is minimally affected.
Recurrent AGN activity, with associated departures from photoionization
equilibrium, can have a stronger impact within $r_i$, but it does not
substantially increase the typical influence radius, so it still
has little effect on the flux decrement close to galaxies.
The predicted dependence of $\meand$ on $\Delta\theta$ {\it is}
sensitive to redshift measurement errors, with rms errors
$\sigma_{\Delta V} \ga 150\kms$ significantly depressing the
absorption signature.

There is significant overlap between our study of the conditional flux
PDF and the recent paper by \cite{croft02a}, who independently
investigate the conditional mean decrement statistic and other
measures of LBG-\lya\ forest correlations, using similar numerical
techniques.  
As discussed in \S\ref{sec:cpdf},
\cite{croft02a} and ASSP define the conditional mean
decrement in a slightly different way from us, averaging over all
galaxy-pixel pairs in bins of redshift-space separation $\Delta_r$
instead of considering only pixels at the galaxy redshift.  Averaging
in redshift-space separation bins instead of angular separation bins
at $\Delta V=0$ (like the ``differential'' points of
Figures~\ref{fig:mean} and~\ref{fig:zerror}) has a small but
noticeable impact, yielding slightly less absorption at a given
comoving distance. 
To facilitate comparison to the
\cite{croft02a} and ASSP results, we have calculated their version of
the mean decrement statistic for the 150 highest SFR galaxies in the
L50n144 simulation ($\Sigma=1\dunits$).  Results are shown in
Figure~\ref{fig:croft}, assuming rms redshift determination errors of
0, 150, and $300\kms$ (triangles, squares, and pentagons,
respectively).

\cite{croft02a} adopt slightly different cosmological parameters, with
$\Omega_m=0.3$, $\sigma_8=0.9$ instead of $\Omega_m=0.4$,
$\sigma_8=0.8$.  Relative to our L50n144 run, their impressive,
$2\times 300^3$ particle simulation has a volume 3.3 times smaller but
mass resolution 30 times higher.  Solid and dotted curves in the left
hand panel of Figure~\ref{fig:croft} show their results for galaxies
with $M_b>10^{10}M_\odot$ and $M_b>10^{11}M_\odot$ (Croft et al.,
fig. 8), kindly provided by R.\ Croft.  Given the differences in
cosmological and numerical parameters, the two independent
calculations agree remarkably well.  The comoving space densities of
the two Croft et al.\ samples, 0.028 and 0.0008 $h^3\,{\rm Mpc}^{-3}$,
correspond to $\Sigma=24\dunits$ and $\Sigma=0.67\dunits$ for our
cosmology, so they are comparable to our $\Sigma=48\dunits$ and
$\Sigma=1\dunits$ samples, respectively.  As shown in
Figure~\ref{fig:mean}, we find no significant difference in the
conditional mean decrement for these two samples, while Croft et al.\
find stronger absorption for the more massive galaxies.  Since their
$M_b>10^{11} M_\odot$ sample comprises only 30 galaxies, this
difference in the results does not seem too serious at present.  Croft
et al.\ also consider higher density samples, with mass resolution
thresholds below the resolution limit of our L22n128 simulation, and
they find a continuing though modest trend of decreasing conditional
mean decrement with decreasing galaxy mass threshold.  In agreement
with Croft et al., we find that redshift determination errors of
$\sigma_{\Delta V} \sim 150-300\kms$ have a substantial impact on the
predicted trend of $\meand$ with galaxy separation, and that galaxy
photoionization does not have a large impact.

ASSP have investigated LBG-\lya\ forest correlations
observationally, carrying out an LBG spectroscopic survey in fields
around six quasar lines of sight.
Their estimates of the conditional mean decrement, kindly
provided by K.\ Adelberger, are shown by the filled points
in Figure~\ref{fig:croft}.  These are somewhat different from
the points shown by \cite{croft02a} because of changes to ASSP's
analysis procedures, in particular a scaling of optical depths
with redshift that should make the results more directly comparable
to predictions like the ones presented here.
The predicted and observed trends agree reasonably well down to
comoving separations $\sim 3\hmpc$, with a smooth increase of
$\sim 0.1$ in $\meand$.  However, ASSP's trend
flattens towards smaller separations, while our predicted trend
rises and steepens.  Most strikingly, ASSP find a {\it decrease}
in $\meand$ at separations $r < 1\hmpc$,
where the predicted absorption is strongest.
If the rms redshift errors are $\sigma_{\Delta V} \sim 300\kms$ rather
than the $\sim 150\kms$ that ASSP estimate, then the discrepancy
between the predictions and observations is reduced, but the two
data points at $r < 1\hmpc$ remain in serious conflict.  

ASSP suggest that the flattening and eventual decline of the mean
decrement in close proximity to LBGs is a signature of galaxy feedback
on the surrounding IGM, perhaps caused by the galactic scale winds
suggested by the nebular line profiles of LBGs
\citep{pettini01,pettini02}.  \cite{croft02a} consider a variety of
galaxy feedback models and show that they can roughly reproduce the
ASSP mean decrement results if $\sim 10\%$ of the available supernova
energy is converted to kinetic energy of a coherent, galactic scale
outflow that sweeps up all of the IGM in its path.  While this wind
model is {\it ad hoc} and idealized, it does show that it is
energetically possible for supernova feedback to influence the
conditional mean decrement out distances of a few $\hmpc$ (comoving).
The range of the effect is larger than that of individual wind bubbles
because the galaxies themselves are correlated, and a line of sight
$\sim 2\hmpc$ from one galaxy may pass through the wind-blown bubble
of another.  
We have shown (in agreement with \citealt{croft02a} and ASSP) that 
galaxy photoionization cannot explain the observed downturn in
absorption, and we have shown that this remains the case even
if one considers departures from ionization equilibrium that could
be caused by recurrent AGN activity.

The results of \S\ref{sec:gpi} (Figure~\ref{fig:dlos} in particular)
have an important implication for wind models: peculiar velocities
allow gas at comoving distances $1-2\hmpc$ to produce strong absorption
at redshift-space separations $0-0.5\hmpc$, so the required wind
radius $R_w$ is larger than one would first guess from the scale
of the observed downturn.  Figure~\ref{fig:croftalt} demonstrates this
point explicitly, using the $\Sigma=3\dunits$ sample of the L22n128
sample (the top 40 galaxies in the box).  Squares, pentagons, and hexagons
show the conditional mean decrement computed after eliminating all
neutral hydrogen within a comoving radius $R_w=0.75$, $1.0$, or $1.5\hmpc$
around each of the target galaxies (see \citealt{kollmeier02} for
further discussion).
Only the $R_w=1.5\hmpc$ case produces
a downturn comparable to the observed one, and even it does not 
reproduce ASSP's innermost data point (as we discuss further below).
Stars show a case in which we put winds around all 641 resolved galaxies
in the box and scale the wind volume in proportion to each galaxy's
baryon mass, $R_w\propto M_b^{1/3}$.  This model has $R_w \approx 1\hmpc$
for the 40th-ranked galaxy in the box and a maximum $R_w \approx 2\hmpc$.
The impact is similar to, though slightly weaker than,
that of the constant-radius $R_w=1.5\hmpc$ model.
All of these models are, of course, highly simplified, but they show
that galactic winds must efficiently eliminate neutral hydrogen out
to large distances to explain ASSP's results.
The energetic requirements for such winds are not impossible to meet
(see \citealt{croft02a}), but they do require sustained outflow speeds 
of several hundred $\kms$ for $t \ga 1$ Gyr and effective entrainment
of surrounding gas.
Winds with large filling factor can have a noticeable impact on the
flux power spectrum of the \lya\ forest, though for $R_w \leq 1\hmpc$
the effect is small \citep{weinberg03}.

The innermost ASSP data point is based on only three galaxy-QSO pairs,
and the error estimation procedure is somewhat {\it ad hoc} (see ASSP),
so it is natural to ask whether the discrepancy with the data is
statistically significant.  To address this point, we randomly selected
3-tuples of simulated galaxy-sightline pairs with 
$\Delta\theta < 0.4'$ and computed the mean decrement in the redshift
separation bin $\Delta_r = 0-0.5\hmpc$, repeating the experiment 500 times.
Figure~\ref{fig:monte} shows the cumulative distribution of these
3-tuple mean decrements, for our standard model with no redshift measurement
errors (solid line), for redshift errors $\sigma_{\Delta V}=300\kms$
(dashed line), and for the $R_w=1.5\hmpc$ ``wind'' model from
Figure~\ref{fig:croftalt}.
Vertical lines mark the unconditional mean flux decrement,
$\meand=0.36$, and the ASSP measurement for this separation, $\meand=0.11$.
With $\sigma_{\Delta V}=300\kms$, the 3-tuple mean decrement is below $0.36$
only $\sim 5\%$ of the time, and none of our 500 trials have a mean
decrement as low as the ASSP value.  Even for the $R_w=1.5\hmpc$ wind
model, only $\sim 5\%$ of the trials have a 3-tuple mean decrement as
low as ASSP's.  We did not incorporate redshift errors in the 
wind model, but (in contrast to the no-wind case) they tend to 
raise the predicted mean decrement by moving the measurement point
further from the influence of the galaxy.
We have only 150 simulated galaxies contributing to the solid and dotted
curves and only 40 contributing to the dashed curve, so our 3-tuples are
not independent and we may therefore underestimate the incidence of
rare, low absorption cases.  Nonetheless, it appears that the ASSP
results cannot easily be explained away by appealing to small number
statistics.

\begin{figure*}
\centerline{
\epsfxsize=4.5truein
\epsfbox[37 66 500 700]{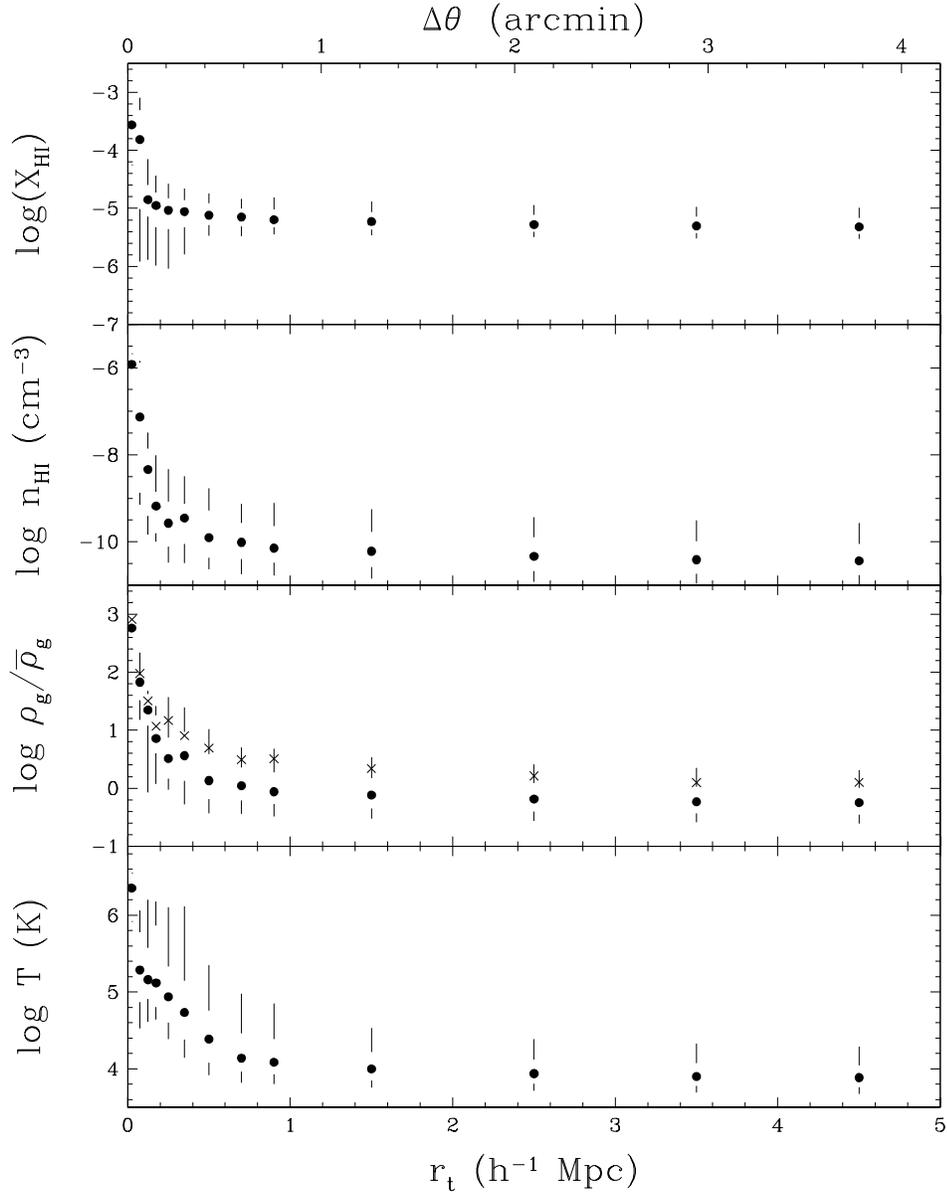}
}
\caption{
Neutral hydrogen fraction, neutral hydrogen density,
gas overdensity, and gas temperature (top to bottom)
as a function of comoving transverse
separation from galaxies, computed using all resolved galaxies in the
L22n128 simulation.  Filled circles show median values for pixels
at the galaxy redshift and this transverse separation, while the inner
and outer edges of the vertical segments mark the 25--75\% and 10--90\%
ranges, respectively.  In the gas density panel, $\times$'s show the
mean overdensity, which is always higher than the median because
of the skewed distribution of $\rho_g/\bar{\rho_g}$.  For our
cosmology, a transverse separation of $1\hmpc$ corresponds to $0.843'$.
Scales along the top and bottom axes are marked in arc-minutes
and comoving $\hmpc$, respectively.
}
\label{fig:profiles}
\end{figure*}

\begin{figure*}
\centerline{
\epsfxsize=4.5truein
\epsfbox[57 221 564 508]{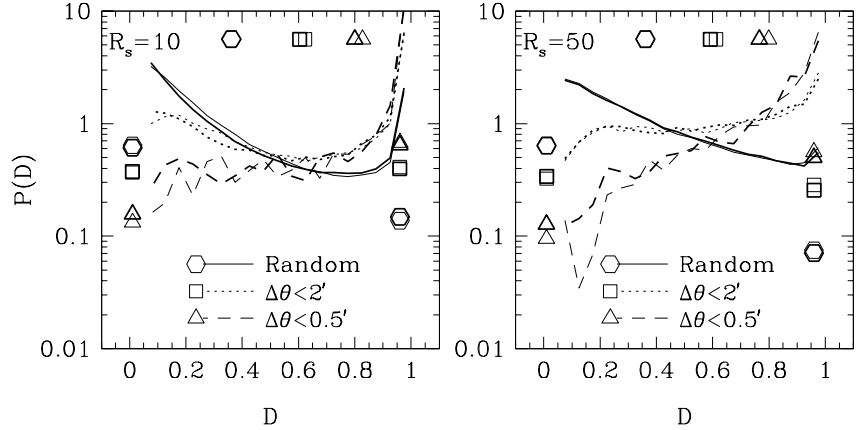}
}
\caption{
Mass resolution effects on the conditional and unconditional flux
PDFs, for spectra smoothed over $10\kms$ (left) or $50\kms$ (right).
Heavy curves and symbols are computed for galaxies with baryonic mass
$M > 512\msph$ in the L11n128 simulation, while light curves and
symbols are computed for galaxies with $M>64\msph$ in the L11n64
simulation.  If there were no resolution effects at all on the galaxy
populations or the spectra, then the results would agree perfectly.
Solid curves show the unconditional PDF, computed using all pixels
along random lines of sight through the simulation box.
Dotted and dashed curves show $P(D)$ for pixels at the same redshift
as a galaxy with transverse separation $\Delta\theta < 2'$ or
$\Delta\theta < 0.5'$, respectively.  Symbols along the top, right,
and left axes show, respectively, the mean flux decrement,
the fraction of ``saturated'' pixels (with $D>0.9$), and the
fraction of ``transparent'' pixels (with $D<0.55$), computed
from the corresponding PDF.
}
\label{fig:resolution}
\end{figure*}

\begin{figure*}
\centerline{
\epsfxsize=4.5truein
\epsfbox[57 221 564 508]{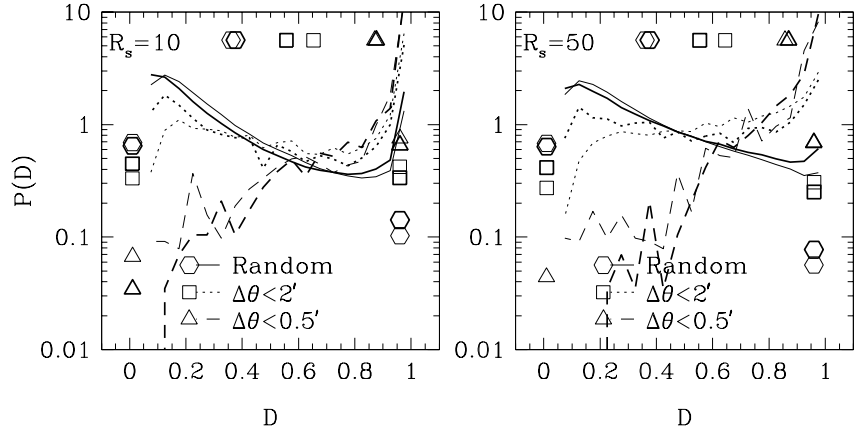}
}
\caption{
Comparison of the unconditional and conditional flux PDFs from the L22n128
and L50n144 simulations, in the same format as Fig.~\ref{fig:resolution}.
Heavy curves and symbols show results using the 37 galaxies with
$M>512\msph$ in L22n128, and light curves and symbols show results
using the 450 galaxies with $M>64\msph$ (the same physical mass threshold)
in L50n144.  Differences can arise from systematic effects of resolution,
box size, or random differences in the structures present in the
two simulations.  In the right hand panel, the heavy triangle representing
the saturated fraction of L22n128 is superposed on the light triangle,
and the heavy triangle representing the transparent fraction is off the
bottom of the plot.
}
\label{fig:resolution2}
\end{figure*}

\begin{figure*}
\centerline{
\epsfxsize=4.5truein
\epsfbox[51 216 562 724]{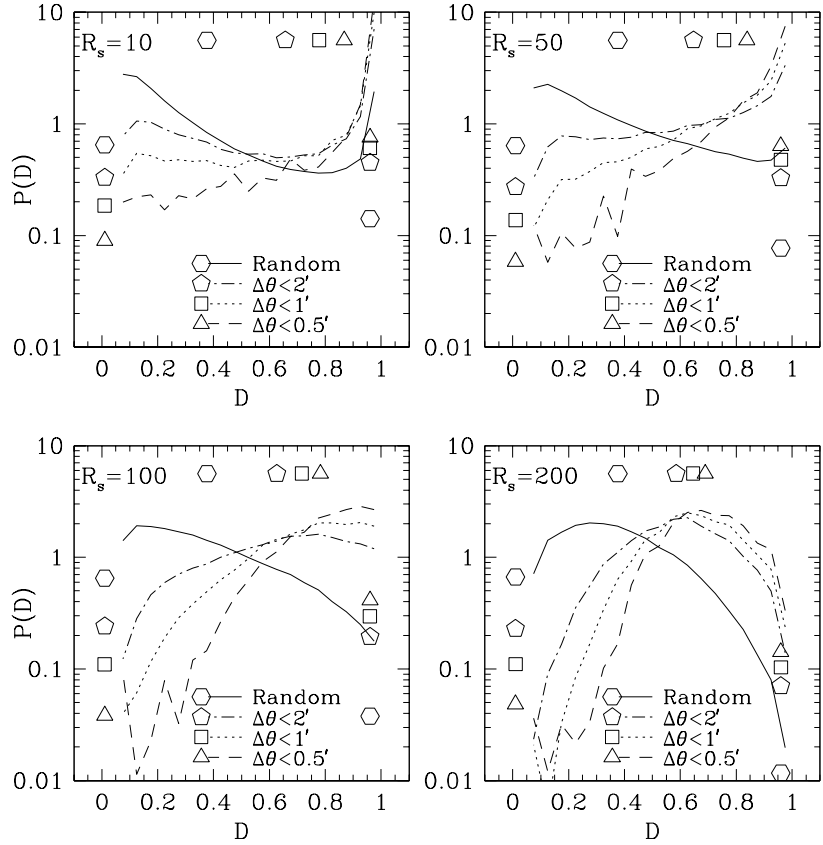}
}
\caption{
The unconditional and conditional flux PDFs computed using all
resolved galaxies ($M>64\msph$) in the L22n128 simulation.
The four panels show results for spectral smoothing lengths
of 10, 50, 100, and $200\kms$ (Gaussian $1\sigma$ width), respectively.
The format is similar to Figs.~\ref{fig:resolution} and~\ref{fig:resolution2}:
solid curves show $P(D)$ for all pixels in random lines of sight,
while dot-dashed, dotted, and dashed curves show $P(D)$ for pixels
at the same redshift as galaxies with transverse separations
$\Delta\theta < 2'$, $1'$, and $0.5'$, respectively.
Symbols along the top, right, and left axes show the mean
decrement, saturated fraction, and transparent fraction.
}
\label{fig:cpdf}
\end{figure*}

\begin{figure*}
\centerline{
\epsfxsize=4.5truein
\epsfbox[57 226 562 724]{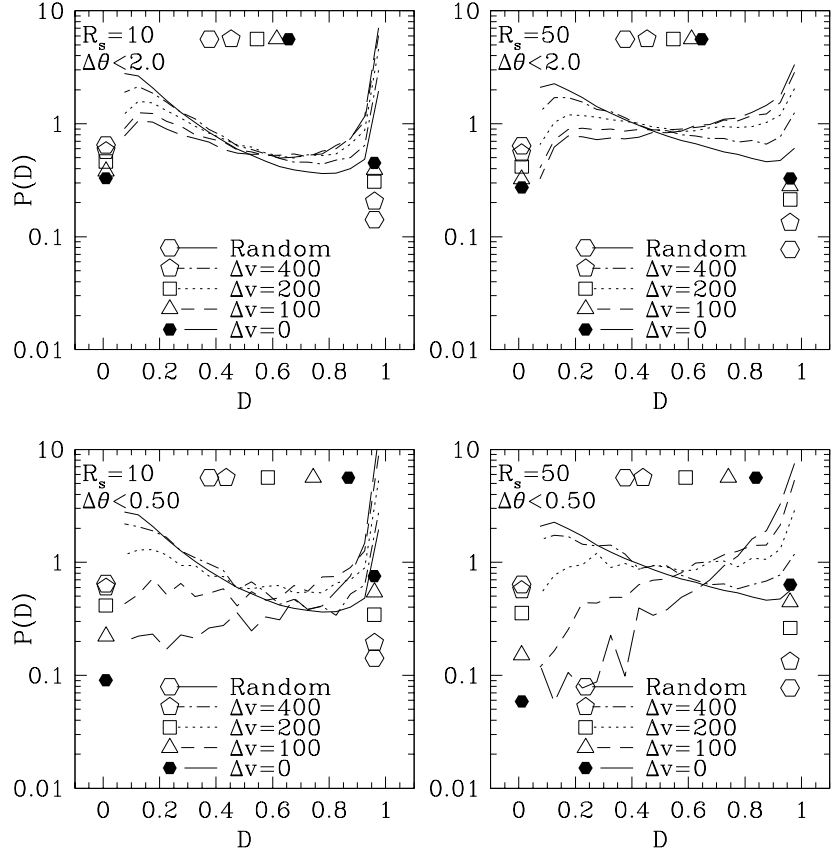}
}
\caption{
Dependence of the conditional flux PDF on velocity offset $\Delta V$,
computed using all resolved galaxies in the L22n128 simulation.
Top and bottom panels show angular proximity conditions
$\Delta\theta < 2'$ and $\Delta\theta < 0.5'$, respectively,
with spectral smoothing of $10\kms$ (left) or $50\kms$ (right).
In each panel, solid curves show the unconditional flux PDF, and
long-dashed curves show the conditional PDF for pixels at the same redshift
as a galaxy, as in Fig.~\ref{fig:cpdf}.
Dashed, dotted, and dot-dashed curves show conditional PDFs in
pixels offset from the galaxy systemic velocity by $|\Delta V|=100$,
200, or $400\kms$, respectively.  Symbols show the corresponding
mean decrements, saturated fractions, and transparent fractions.
}
\label{fig:deltav}
\end{figure*}

\begin{figure*}
\centerline{
\epsfxsize=4.5truein
\epsfbox[57 221 564 508]{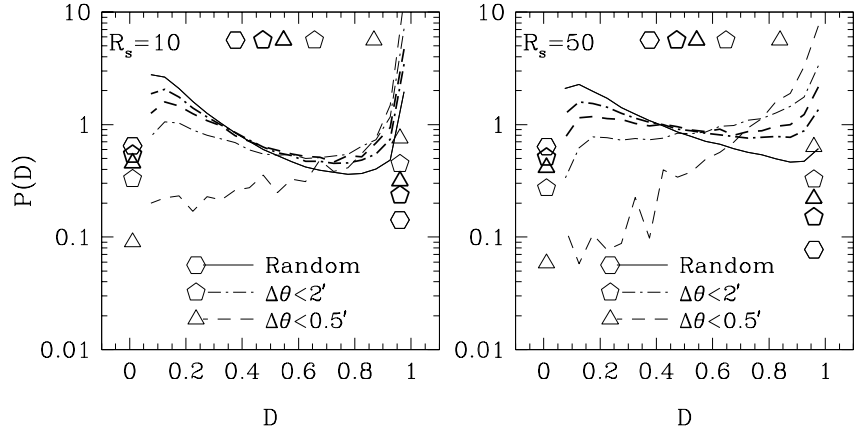}
}
\caption{
The impact of galaxy bias on the conditional flux PDF, for spectral
smoothing lengths of $10\kms$ (left) and $50\kms$ (right).
Light solid, dot-dashed, and dashed curves, repeated from the
corresponding panels of Fig.~\ref{fig:cpdf}, show the unconditional
flux PDF and the conditional flux PDFs for $\Delta\theta < 2'$ and
$\Delta\theta < 0.5'$ about resolved galaxies in the L22n128 simulation.
Heavy curves show the corresponding conditional flux PDFs using
randomly selected dark matter particles in place of galaxies.
Light and heavy symbols are based on the corresponding flux PDFs.
While the conditional flux PDF about randomly selected mass elements
differs from the unconditional flux PDF (which is effectively
volume weighted), the deviations of the galaxy conditional PDFs are
much stronger, a clear signature of the preferential formation
of high redshift galaxies in overdense large scale environments.
}
\label{fig:bias}
\end{figure*}

\begin{figure*}
\centerline{
\epsfxsize=4.5truein
\epsfbox[58 212 564 727]{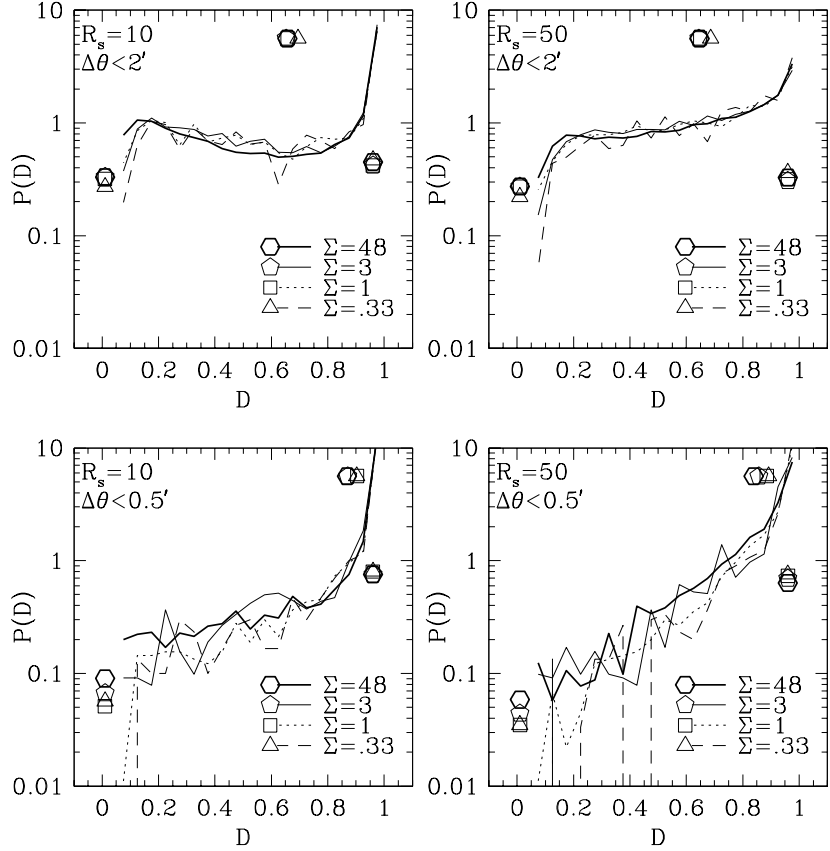}
}
\caption{
Dependence of the conditional flux PDF on galaxy mass and SFR.
In each panel, heavy solid lines show the conditional PDF
based on all resolved galaxies in the L22n128 simulation, as in
Fig.~\ref{fig:cpdf}.  The baryonic mass threshold is
$M \geq 6.7\times 10^9M_\odot$, and the space density of
the population is $\Sigma=48\dunits$.
Light solid lines show results for the resolved galaxy
population of L50n144, with a mass threshold of $5.4\times 10^{10}M_\odot$
and a space density of $\Sigma=3\dunits$.  Dotted and dashed lines
are for galaxies in L50n144 with
${\rm SFR}>108 M_\odot {\rm yr}^{-1}$ and
${\rm SFR}>180 M_\odot {\rm yr}^{-1}$, respectively,
with population space densities $\Sigma=1\dunits$ and $\Sigma=0.33\dunits$.
}
\label{fig:luminosity}
\end{figure*}

\begin{figure*}
\centerline{
\epsfxsize=4.5truein
\epsfbox[51 466 570 729]{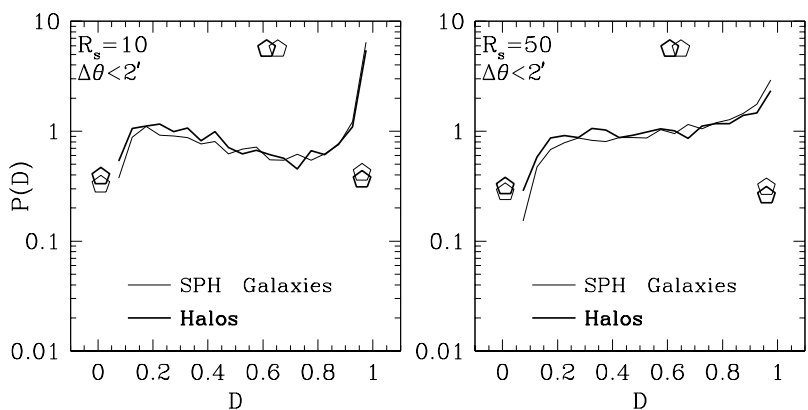}
}
\caption{
Comparison of halo and galaxy conditional flux PDFs.
Light lines and symbols, repeated from the corresponding
panels of Fig.~\ref{fig:luminosity}, show results for the
resolved galaxies in L50n144, with $\Sigma=3\dunits$.
Heavy lines and symbols show results when these galaxies are
replaced by friends-of-friends halos identified from the simulation's
dark matter particle distribution.  Halos are selected above
a (dark matter) mass threshold of $5.7\times 10^{11} M_\odot$,
yielding 450 halos in the simulation volume and $\Sigma=3\dunits.$
}
\label{fig:halo}
\end{figure*}

\begin{figure*}
\centerline{
\epsfxsize=4.5truein
\epsfbox[51 466 570 729]{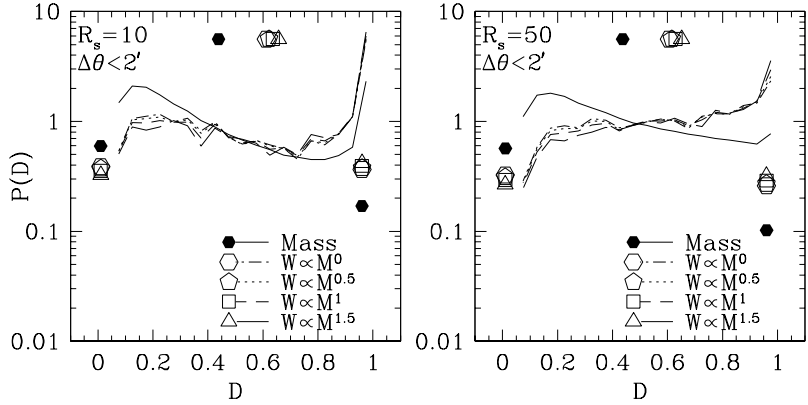}
}
\caption{
Influence of halo weighting on the conditional flux PDF.
Dot-dashed lines, repeated from Fig.~\ref{fig:halo}, show results
in which each halo above the mass threshold contributes equally.
Solid lines and filled symbols show results for randomly selected
dark matter particles.  Other lines show results in which the contribution
of each halo-pixel pair to the conditional PDF is weighted by $M^\alpha$,
with $\alpha=0.5$ (dotted), $\alpha=1$ (short-dashed), or
$\alpha=1.5$ (long-dashed).  While the overall bias of halos makes the
conditional PDF of halos substantially different from that of the mass,
stronger weighting of high mass halos has only a small impact
on this statistic.
}
\label{fig:hod}
\end{figure*}

\begin{figure*}
\centerline{
\epsfxsize=2.5truein
\epsfbox[73 450 325 729]{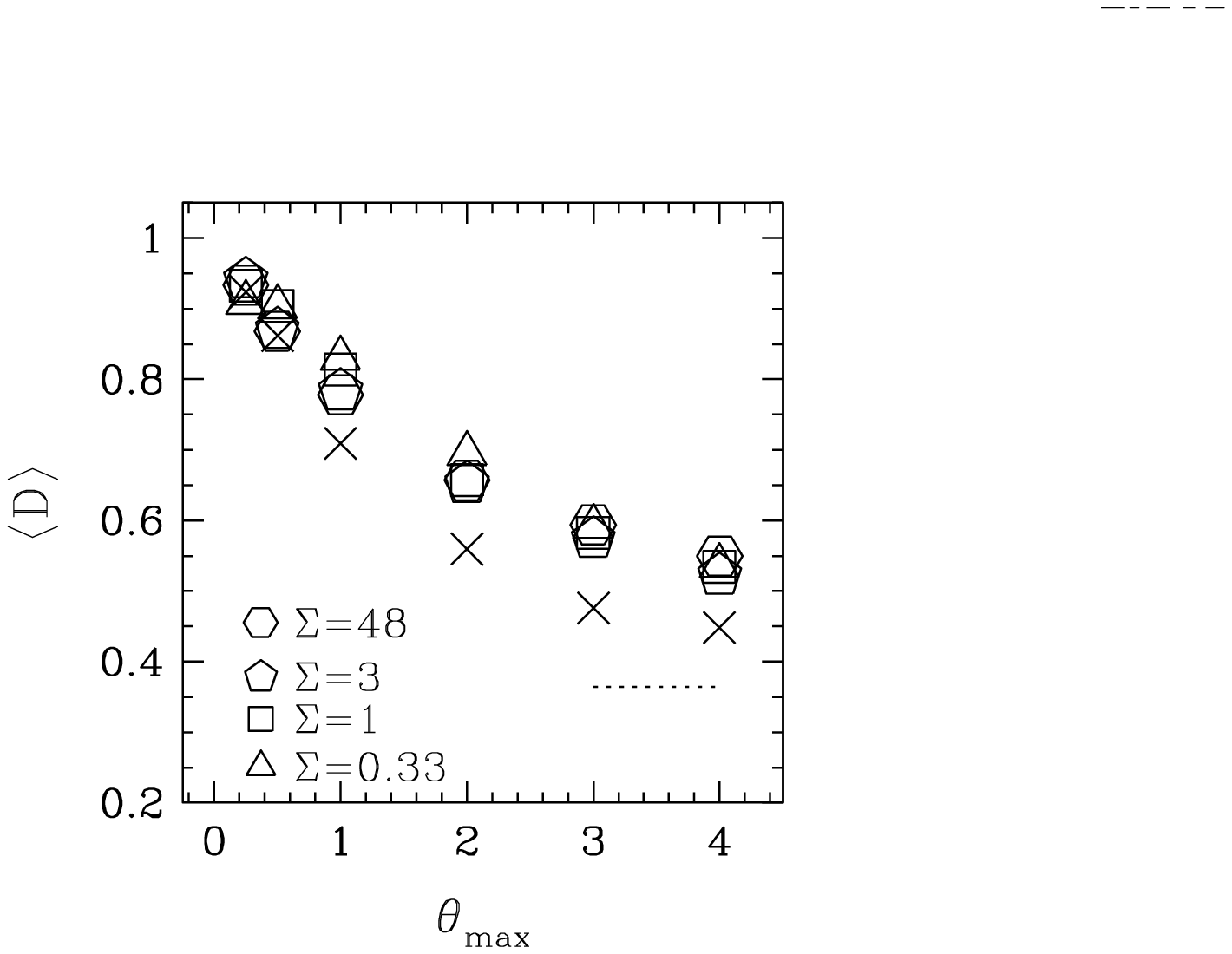}
}
\caption{
The conditional mean flux decrement, for spectral smoothing of $10\kms$.
Open symbols show the mean decrement of pixels at the redshift of
a galaxy with angular separation $\Delta\theta < \tmax$ arc-minutes.
Results are shown for the resolved galaxy population of L22n128
($\Sigma=48\dunits$), the resolved galaxy population of L50n144
($\Sigma=3\dunits$), and SFR-thresholded samples of the L50n144 population
with $\Sigma=1\dunits$ and $\Sigma=0.33\dunits$.
The $\times$'s show a ``differential'' form of this statistic,
with $\meand$ computed for separations in the range
$0.8 \tmax \leq \Delta\theta < 1.25\tmax$, for the $\Sigma=3\dunits$
case only.
The horizontal dotted line segment marks the unconditional
mean decrement, $\meand = 0.36$.  For our cosmology, $1'$ corresponds
to $1.19\hmpc$ (comoving) and $H r_t = 152\kms$ at $z=3$.
}
\label{fig:mean}
\end{figure*}

\begin{figure*}
\centerline{
\epsfxsize=2.5truein
\epsfbox[73 450 325 729]{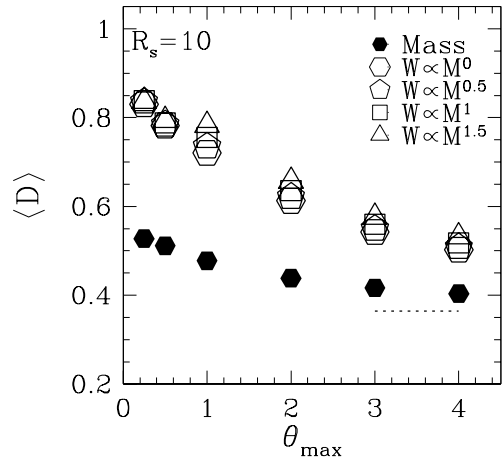}
}
\caption{
The conditional mean flux decrement (cumulative form) for randomly
selected dark matter particles (filled symbols) and for dark matter
halos with different relative weightings (open symbols, with weights
as marked in the legend), computed from the L50n144 simulation.
The dotted horizontal line segment
shows the unconditional mean decrement $\meand = 0.36$.
}
\label{fig:mean2}
\end{figure*}

\begin{figure*}
\centerline{
\epsfxsize=4.5truein
\epsfbox[54 480 571 734]{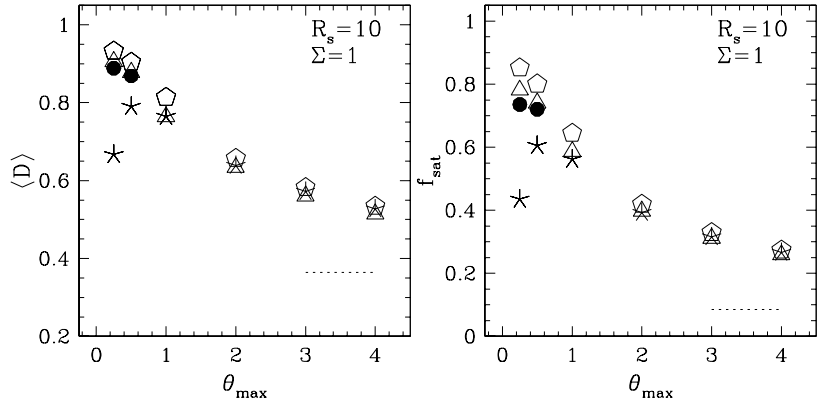}
}
\caption{
Potential impact of galaxy photoionization on the conditional
mean flux decrement (left) and the saturated pixel fraction (right),
for the $\Sigma=1\dunits$ sample of the L50n144 simulation and
a spectral smoothing of $10\kms$.
We assume that these galaxies produce a fraction $f=0.5$ of the
UV background at $z=3$, with each galaxy's photoionizing flux
proportional to its SFR.  Pentagons show results in the absence
of galaxy photoionization, asterisks show the results of the
approximate calculation described in the text,
and triangles show
the results of a complete calculation with the UV sources embedded
in the simulation before spectra are extracted.
The complete calculation yields a much weaker effect because peculiar
velocities allow relatively distant gas to produce absorption at the galaxy
redshift (see Fig.~\ref{fig:dlos}).
Filled circles show results for the recurrent AGN model
described in the text, for $\tmax=0.25'$ and $\tmax=0.5'$ only.
Dotted horizontal segments show the mean decrement $\meand = 0.36$
or saturated fraction $\fsat = 0.086$ for the unconditional flux PDF.
}
\label{fig:gpi}
\end{figure*}

\begin{figure*}
\centerline{
\epsfxsize=6.0truein
\epsfbox[25 480 530 675]{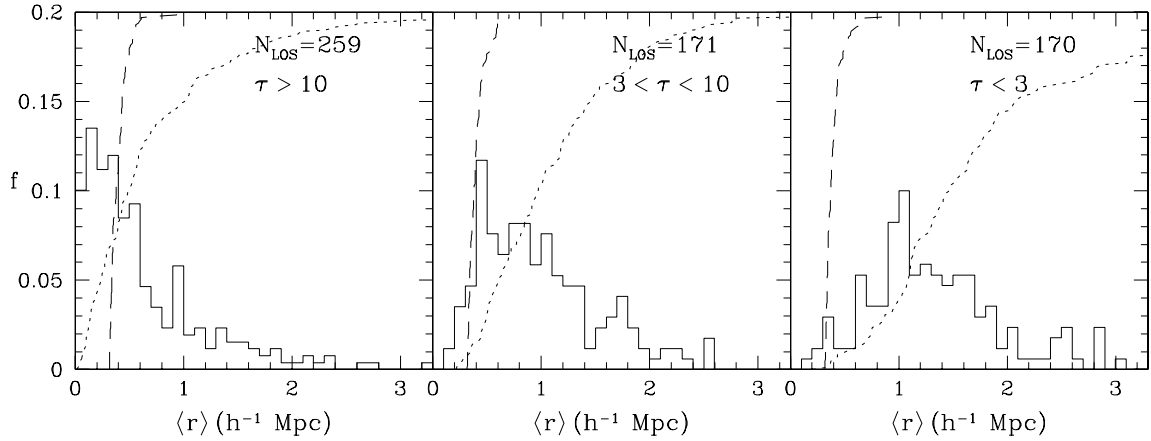}
}
\caption{
Distribution of the mean line-of-sight distance, weighted by optical depth,
of gas producing absorption within $20\kms$ of the galaxy redshift,
calculated
for lines of sight within $0.5'$ of galaxies in the $\Sigma=1\dunits$
sample of the L50n144 simulation, assuming
a uniform ionizing background.  Left, middle, and right panels show
three different ranges of the optical depth at the galaxy redshift,
with the number of lines of sight as indicated.
In each panel, histograms show the differential distribution and dotted
lines the cumulative distribution (on a scale where the top of the
panel is unity).  Dashed lines show the cumulative distribution of
photoionization influence radii $r_i$ for the 150 galaxies.
Units are comoving $\hmpc$.
}
\label{fig:dlos}
\end{figure*}

\begin{figure*}
\centerline{
\epsfxsize=4.5truein
\epsfbox[57 226 562 724]{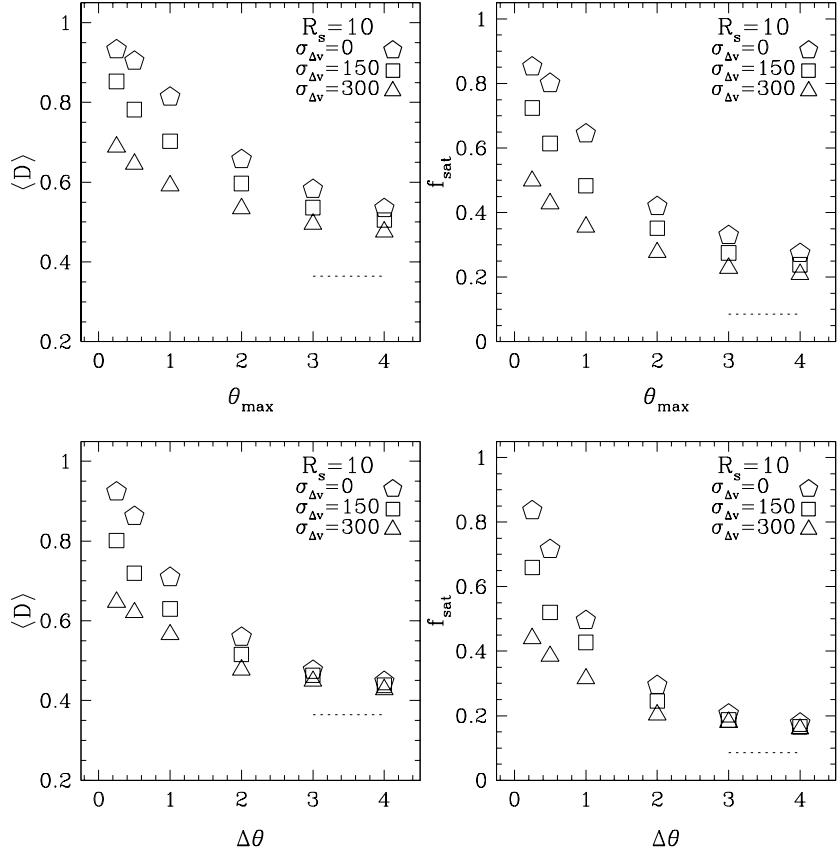}
}
\caption{
Impact of redshift determination errors on the conditional mean
decrement (left) and saturated pixel fraction (right).
Upper panels show our standard, cumulative form of these
statistics, computed for all galaxy-pixel pairs with
angular separation $\Delta\theta<\tmax$,
and lower panels show the differential form, computed for pairs with
$0.8\tmax < \Delta\theta < 1.25\tmax$.
In each panel, pentagons
show results for the $\Sigma=1\dunits$ sample of the L50n144
galaxy population in the absence of redshift errors.
Squares and triangles
show the corresponding results when estimated galaxy redshifts are drawn
from a Gaussian distribution of $1\sigma$ dispersion $150\kms$
or $300\kms$, respectively, centered on the true galaxy redshifts.
}
\label{fig:zerror}
\end{figure*}

\begin{figure*}
\centerline{
\epsfxsize=4.5truein
\epsfbox[50 210 550 720]{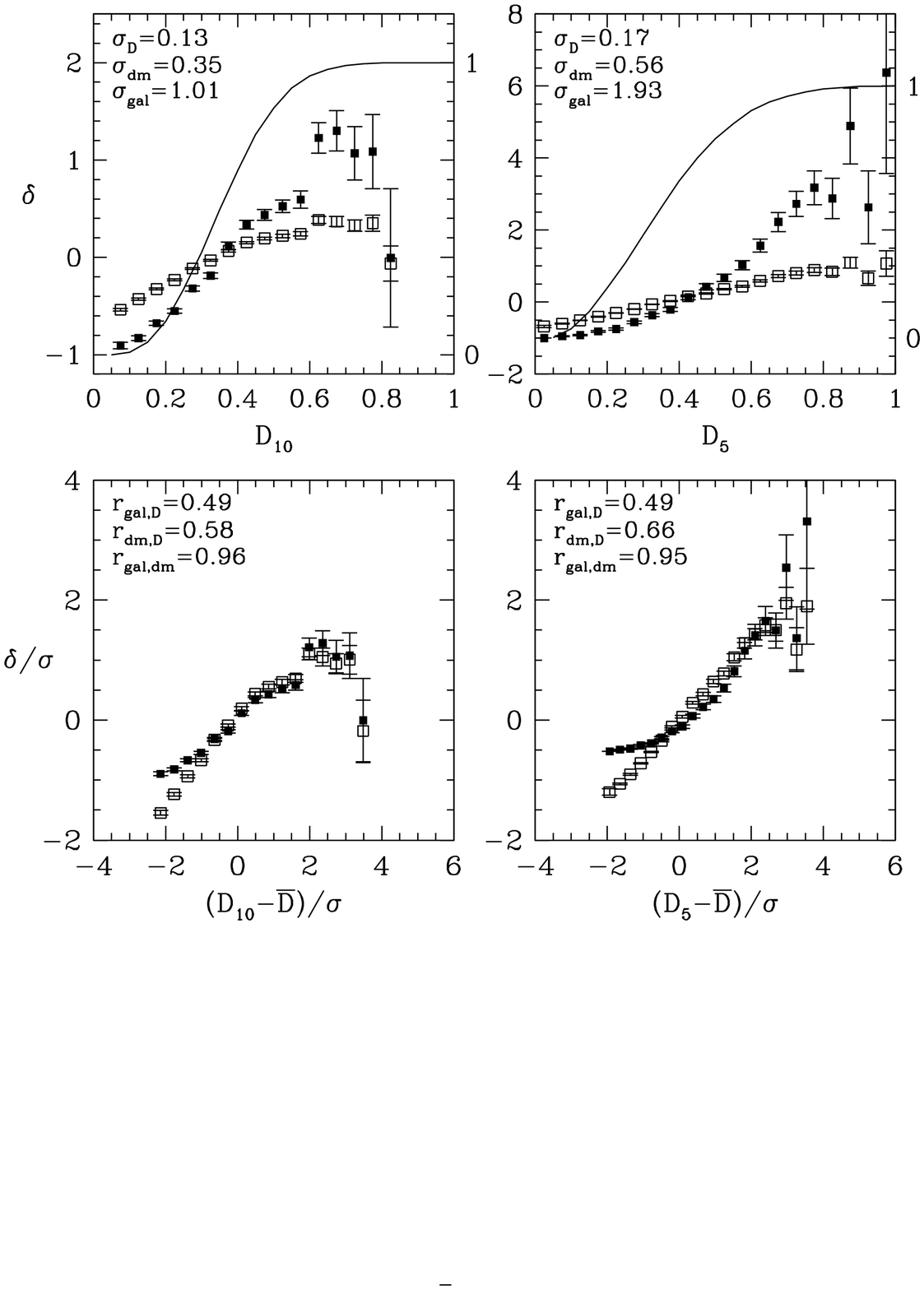}
}
\caption{
Large scale correlations of galaxy and dark matter overdensity
with \lya\ flux decrement, in cubes of comoving size $10\hmpc$
(left) and $5\hmpc$ (right).  The mean decrement along a segment
of \lya\ forest spectrum traversing the center of the cube is
denoted $D_{10}$ or $D_5$, and filled (open) squares in the top
panel show the mean value of the galaxy (dark matter) density
contrast in cubes with a specified value of $D_{10}$ or $D_5$.
Plots in the lower panels are normalized by subtracting the
mean and dividing by the dispersion of the corresponding variable.
Curves in the upper panels show the cumulative distribution of
$D_{10}$ and $D_{5}$, with values zero and one marked on the right axis.
Upper panels list dispersions of the decrement, dark matter contrast,
and galaxy contrast (with shot noise subtracted),
and lower panels list cross-correlation coefficients.
All error bars show $1\sigma$ uncertainty in the mean; the
dispersion of $\delta$ values in a given bin of $D_{10}$
or $D_5$ is much larger.
}
\label{fig:cube}
\end{figure*}

\begin{figure*}
\centerline{
\epsfxsize=4.5truein
\epsfbox[50 210 550 720]{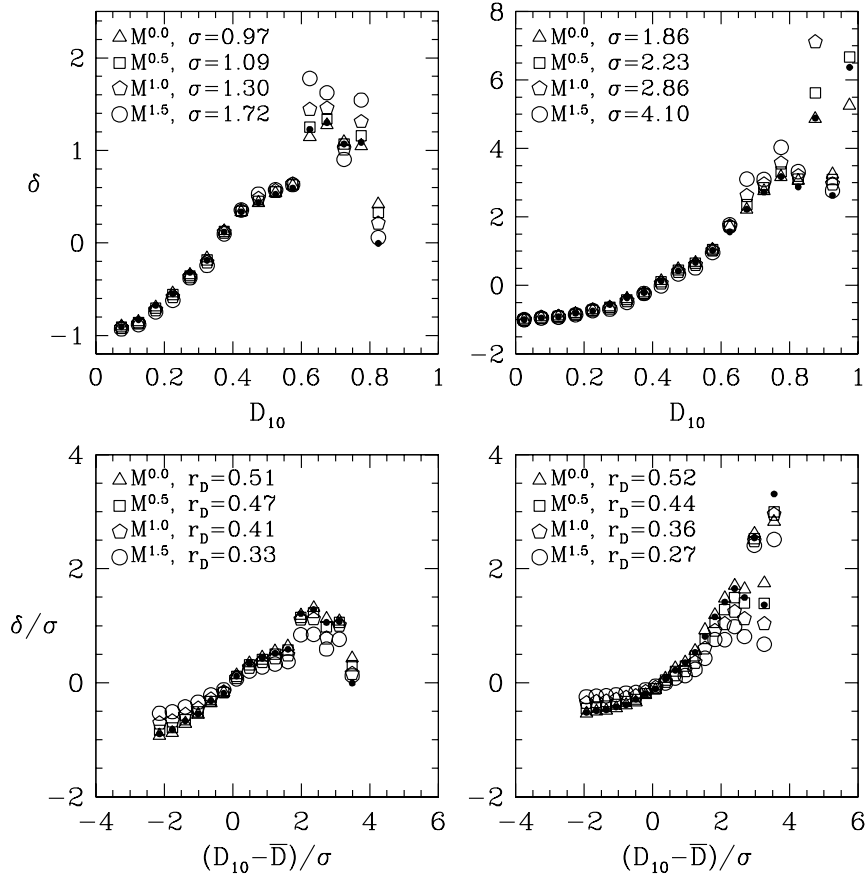}
}
\caption{Similar to Fig.~\ref{fig:cube}, but for dark matter halos.
Triangles show the case in which each of the 450 most massive
halos contributes equally.  Squares, pentagons, and circles
show mass-weighted cases, with the contribution of each halo proportional to
$M^{0.5}$, $M$, and $M^{1.5}$, respectively.  Solid dots show the
results for galaxies, repeated from Fig.~\ref{fig:cube}.
}
\label{fig:cubehod}
\end{figure*}

\begin{figure*}
\centerline{
\epsfxsize=4.5truein
\epsfbox[50 462 315 720]{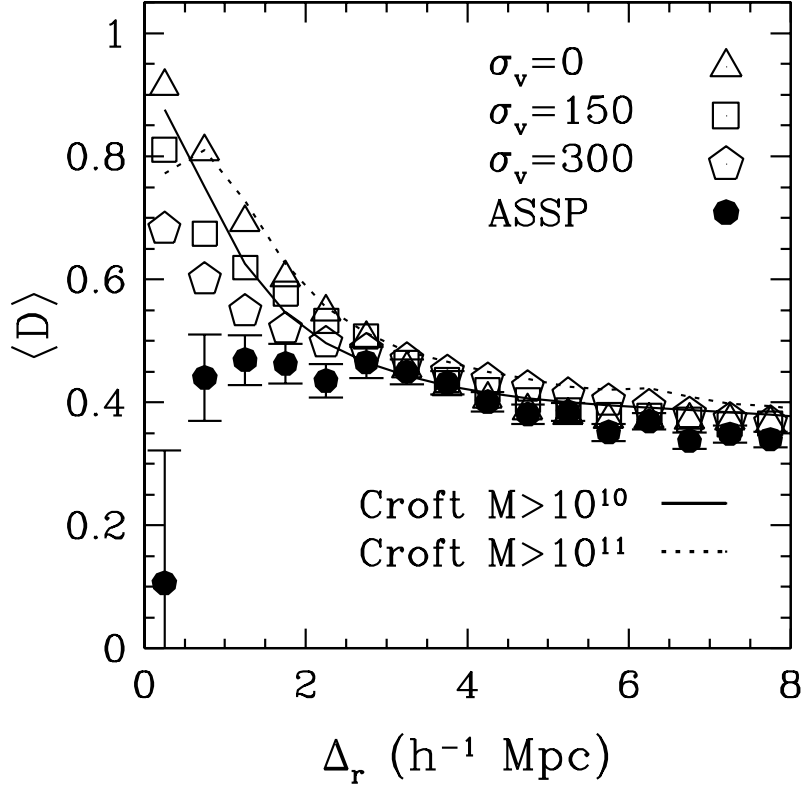}
}
\caption{ The conditional mean flux decrement, now defined by an
average over all galaxy-pixel pairs in bins of comoving redshift-space
separation.  Open symbols show our results for the $\Sigma=1\dunits$
sample of the L50n144 simulation, with spectral smoothing length
$R_s=10\kms$ and no redshift errors (triangles), rms redshift errors
of $150\kms$ (squares) and $300\kms$ (pentagons).  Solid and dotted
curves show Croft et al.'s (2002a) results for galaxies with baryonic
mass $M_b>10^{10} M_\odot$ and $M_b>10^{11}M_\odot$, respectively.
Filled symbols show the ASSP observational results, with approximate
error bars estimated by a Monte Carlo procedure in which galaxy
redshifts are perturbed by $\sim 50\hmpc$ (comoving) to produce a
``random'' sample with the same transverse separations.}
\label{fig:croft}
\end{figure*}

\begin{figure*}
\centerline{
\epsfxsize=4.5truein
\epsfbox[50 462 315 720]{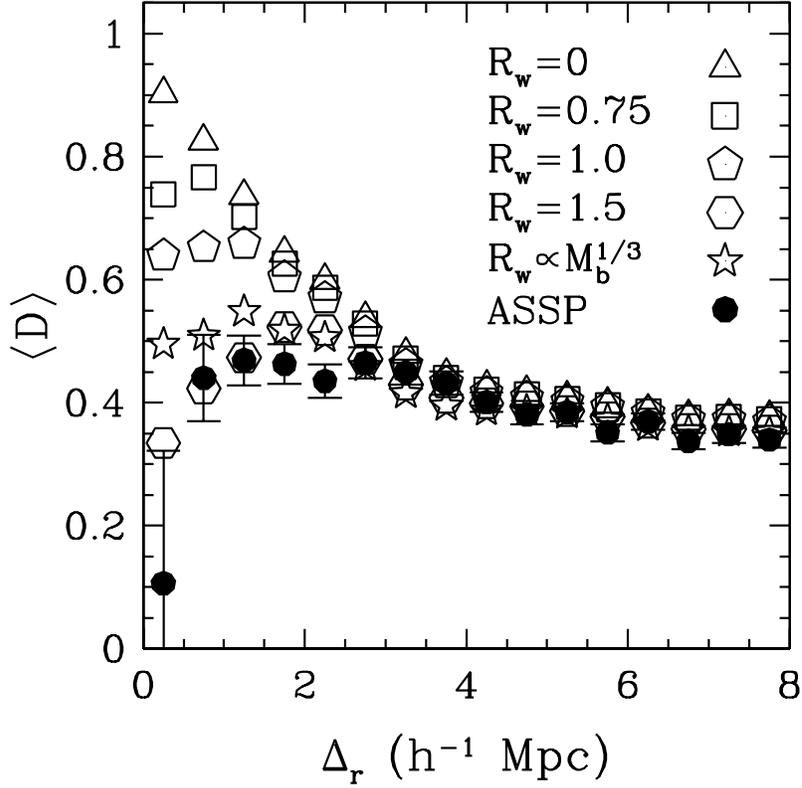}
}
\caption{Effects of simplified ``winds'' on the mean flux
decrement. Open symbols show our results for the $\Sigma=3\dunits$
sample of the L22n128 simulation, with spectral smoothing length
$R_s=10\kms$ and wind radii of 0.75 (squares), 1.0 (pentagons), 1.5
$\hmpc$ (hexagons) and proportional to $M_b^{1/3}$ (stars).  Open
triangles show our results for no winds, and filled symbols show the
ASSP observational results.  }
\label{fig:croftalt}
\end{figure*}

\begin{figure*}
\centerline{
\epsfxsize=4.5truein
\epsfbox[50 470 315 720]{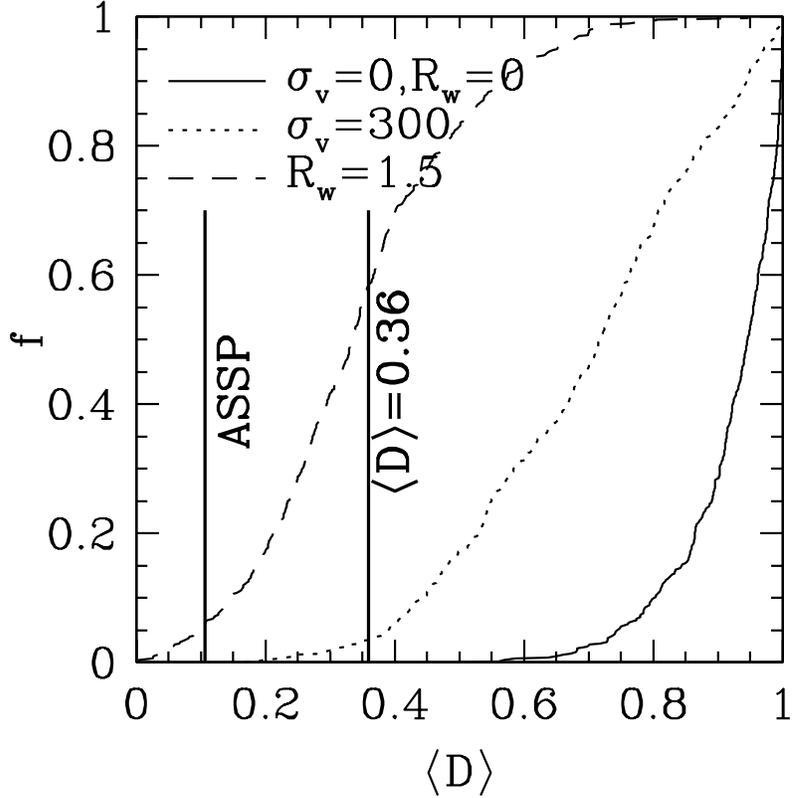}
}
\caption{Statistical significance of the discrepancy with the
innermost ASSP data point, which comes from three galaxy-QSO pairs.
Vertical lines mark the ASSP measured value for $\Delta_r < 0.5\hmpc$,
$\meand=0.11$, and the unconditional mean flux decrement, $\meand=0.36$.
For each model, we randomly select 500 3-tuples of close galaxy-spectrum
pairs and compute the mean decrement for pixels with $\Delta_r<0.5\hmpc$.
Curves show the cumulative distribution of these mean decrement values,
for our standard model with no redshift errors (solid, corresponding
to the triangles in Fig.~\ref{fig:croft}), the standard model with 
$300\kms$ rms redshift errors (dotted, corresponding to pentagons
in Fig.~\ref{fig:croft}), and the $R_w=1.5\hmpc$ wind model with
no redshift errors (dashed, corresponding to the hexagons in
Fig.~\ref{fig:croftalt}).
}
\label{fig:monte}
\end{figure*}

Given the stringent demands on wind models, it is worth considering
alternative explanations.  One possibility is that the IGM at these
distances from galaxies is multi-phase, with processes that the
simulations cannot resolve causing the geometrical cross section of
the neutral phase to drop sharply.  However, the typical physical
conditions at these distances from galaxies are not extreme (see
Figure~\ref{fig:profiles}), so it would be difficult to reconcile such
an explanation with the overall success of simulations in reproducing
(unconditional) statistical properties of the \lya\ forest.  

A final possibility worth considering is that extended \lya\ {\it
emission} from the target galaxies ``fills in'' the \lya\ forest at
the redshift of emission.  Extended \lya\ ``blobs'' have been observed
to be associated with LBG's \citep{steidel00} having angular extents
$\sim 15''$ and AB apparent magnitudes 21.02 and 21.14
in an 80\AA\ \lya\ band.  Cooling radiation
from gas settling into massive galaxies at $z=3$ naturally produces
\lya\ flux of this order, and numerical simulations predict that such blobs
should be present around typical LBGs \citep{fardal01}.  
Thus, a galaxy's \lya\ cooling radiation could plausibly replace the 
absorbed continuum flux of a 21st-magnitude background quasar.
There are two substantial and partly cancelling corrections to 
this estimate.  First, the \lya\ emission extends over 
$\sim 1000\kms$ (16 \AA) rather than the 80\AA\ bandpass used
by \cite{steidel00}, so the flux density at the \lya\ wavelength
is a factor of five higher ($-1.75$ mag).
Second, at a separation $\Delta\theta \sim 15-20"$ from a 
target galaxy, the $\sim 0.8"$ slits used by ASSP should intercept
only $\sim 0.8/2\Delta\theta \sim 0.02-0.03$ of the
galaxy's extended \lya\ flux ($\sim +4$ mag), 
assuming constant surface brightness out to $\Delta\theta$.
The three quasars that contribute to ASSP's innermost data point have $G$-band
AB magnitudes of 20.1, 21.6, and 23.4, so with the $+2.25$ mag net
correction, it appears 
that \lya\ emission could replace the absorbed quasar flux only for
the faintest of the three quasar targets, unless the galaxies in
question are even brighter than the \cite{steidel00} blobs.
Furthermore, a fourth pair involving the $G=17.8$
quasar Q0302-0019 shows no sign of absorption near the galaxy redshift,
and in this case the quasar is clearly too bright for galaxy emission
to compete with it.
(This pair and two others are dropped from ASSP's $\meand$ calculation
because of possible Ly$\beta$ contamination.)
At this point, therefore, the \lya\ emission explanation
seems unlikely, but future observations that have more close
pairs involving bright quasars will be able to test it conclusively.

On the larger scales probed by the $\dgal-\meands$ correlation,
our predictions are in qualitative agreement with the observational
results reported by ASSP and ASPS.  As the Fluctuating Gunn-Peterson
Approximation suggests, higher mean flux decrements tend to arise in
cells of higher mass density, and these in turn have higher average
galaxy overdensity.  The limited size of our largest volume simulation,
a single $50\hmpc$ cube, leaves statistical and systematic uncertainties
in our predictions.  Nonetheless, our results for the $\ddark-\meands$
correlation at $S=10\hmpc$ are similar to those found by \cite{mcdonald02}
using hydro-PM simulations (with $40\hmpc$ cubes but multiple realizations,
and significantly higher mass resolution).
The $\dgal-\meands$ correlation is much steeper than the $\ddark-\meands$
correlation because of the strong bias of the high redshift galaxy
population, and the suppression of galaxy formation in underdense
environments is especially clear.
The ratio of correlation slopes is approximately equal
to the ratio of rms fluctuation amplitudes, $\siggal/\sigdm$, but a
``linear bias'' model does not fully capture the behavior of the
$\dgal-\meands$ correlation, in particular the curvature at low $\meands$
imposed by the constraint that $\dgal \geq -1$.

Our $50\hmpc$ simulation predicts a correlation coefficient
$\rgalD=0.49$ between
galaxy density contrast and averaged flux decrement in $5\hmpc$ or $10\hmpc$
cells.  This is smaller than the value $r=0.68 \pm 0.06$ that
ASPS measure for $S\approx 10\hmpc$, with a $1\sigma$ error bar
derived from the dispersion among five lines of sight.
However, we cannot assign a reliable ``theoretical error bar'' to
our predicted $\rgalD$ without modeling ASPS's procedures in greater
detail.  Even for an ideal, volume-limited sample, there are subtle
issues associated with shot noise subtraction and estimator biases
(Hui \& Sheth, in preparation), and corrections for redshift selection
functions or sample incompleteness could also have a significant
impact on estimates of $r$ and its uncertainty.
Accurate assessment of the discrepancy between predicted and
observed correlation coefficients probably requires larger simulation
volumes, which would allow creation of multiple independent ``surveys''
comparable to the observed one.  Fortunately, we find that the
$\delta-D_S$ correlations for galaxies are similar to those for
halos of the same space density, so it should be possible to
carry out such assessments using large N-body simulations and the
PM or hydro-PM approximation \citep{croft98,gnedin98} for the \lya\ forest.
The absolute trend (i.e., not normalized by standard deviations)
of $\dgal$ with $\meands$ is the best test for the strong bias
of high redshift galaxies predicted by theoretical models.

As the size of LBG spectroscopic samples grows, moderate resolution
spectra of fainter quasars can greatly increase the number of lines
of sight available for measuring LBG-\lya\ forest correlations.
These larger samples will shrink the error bars of the existing
measurements, and they will eventually allow measurement of the
full conditional flux PDF for different subsets of the LBG population.
The potential impact of redshift determination errors can be investigated
empirically by comparing results for galaxies whose redshifts are
estimated by different methods (e.g., emission vs.\ absorption lines).
Models with galactic winds or other forms of feedback can be tested
using the full PDF rather than the mean decrement alone.
Variations in feedback from one galaxy to another should widen the range of
flux decrement values in addition to reducing the mean,
and winds that sweep up the IGM should produce many
lines of sight with essentially zero flux decrement.
Perhaps the best diagnostic of galaxy feedback will be separate
measurements for galaxy subsets with high and low star formation
rates, or older and younger stellar populations, since one would
expect the young/high SFR galaxies to have had more vigorous
activity in the recent past.  Systematic differences in the
activity of different galaxy populations should produce systematic
differences in the corresponding conditional flux PDFs, and our
results show that such differences would be unlikely to arise
from differences in the spatial clustering of these populations.
A full accounting of the relation between LBGs and the \lya\ forest
will test our basic expectations for the way that dark matter
shapes the large scale distribution of high redshift galaxies, and it
will inform our picture of the ways that galaxies can affect the
high redshift IGM.

\acknowledgments

We thank Kurt Adelberger, Rupert Croft, Mark Fardal, Patrick McDonald,
Jordi Miralda-Escud\'e, Max Pettini, Alice Shapley, and Chuck Steidel
for helpful discussions of these issues and of their results.
We also thank Kurt Adelberger and Rupert Croft for providing
results shown in Figure~\ref{fig:croft}.
JAK acknowledges the support of a university fellowship at Ohio State,
and DHW acknowledges the hospitality and support of the
Institut d'Astrophysique de Paris and the French CNRS during
much of this work.  This project was also supported by
NASA grant NAG5-3525, NSF grant AST-0098515, and Hubble Fellowship
Grant HST-HF-01128.01-A from the Space Telescope Science Institute.

\begin{deluxetable}{lccccc}
\tablecolumns{6}
%\tablewidth{10pc}
\scriptsize
\tablecaption{Simulation Parameters\tablenotemark{a}}
\tablehead{
\colhead{Name} & \colhead{$L_{\rm box}$ $(\hmpc)$} & \colhead{$N$} &
\colhead{$\epsilon_{\rm grav}$ $(\hkpc)$} & \colhead{$64 m_{dark}$ $(\msun)$}
 &\colhead{$64 m_{sph}$ $(\msun)$}}
\startdata
$\ba$  & $11.11$ & $64^{3}$  & $3.5$   & $4.9 \times 10^{10}$ & $6.7 \times 10^9$ \\
$\bb$ & $11.11$ & $128^{3}$ & $1.75$ & $6.2 \times 10^{9}$ & $8.3 \times 10^8$
\\
$\bc$ & $22.22$ & $128^{3}$ & $3.5$ &  $4.9 \times 10^{10}$ & $6.7 \times 10^9$
\\
$\bd$ & $50.00$ & $144^{3}$ & $7$  & $4.0 \times 10^{11}$ & $5.4 \times 10^{10}$ \\
\enddata
\tablenotetext{a}{
Length units are comoving.
In all cases, the adopted cosmology is $\Omega_m=0.4$, $\Omega_\Lambda=0.6$,
$h=0.65$, $\Omega_b=0.02h^{-2}$, $n=0.95$, $\sigma_8=0.80$, and the
output redshift is $z=3$.  The angular size, velocity depth, and redshift
depth of the $22.222\hmpc$ cube are $18.74'$, $2844\kms$, and $0.0379$,
and they scale in proportion to $L_{\rm box}$ for the other simulations.
$\epsilon_{\rm grav}$ is the gravitational softening length
(equivalent Plummer softening), setting the approximate spatial resolution.
}
\label{tab:sims}
\end{deluxetable}

\newpage

\clearpage

\clearpage

\clearpage

\clearpage

\clearpage

\clearpage

\clearpage

\clearpage

\clearpage

\clearpage

\clearpage

\clearpage

\clearpage

\clearpage

\clearpage

\clearpage

\clearpage

\clearpage

\clearpage


\begin{thebibliography}{}

\bibitem[Adelberger et al.(1998)]{adelberger98}
Adelberger, K. L., Steidel, C. C., Giavalisco, M., Dickinson, M.,
Pettini, M., \& Kellogg, M. 1998, \apj, 505, 18

\bibitem[Adelberger et al.(2002a)]{adelberger02}
Adelberger, K.L., Steidel, C.C., Shapley, A.E., Pettini, M. 2002a, \apj, 584,45 (ASSP)

\bibitem[Adelberger et al.(2002b)]{adelberger02b}
Adelberger, K.L., Steidel, C.C., Pettini, M., \& Shapley, A. E. 2002b,
in preparation (ASPS)

\bibitem[Aguirre et al.(2001)]{aguirre01}
Aguirre, A., Hernquist, L., Schaye, J., Weinberg, D. H.,
Katz, N., \& Gardner, J. 2001, \apj, 561, 521
% enrichment methods paper

\bibitem[Bagla(1998)]{bagla98}
Bagla, J.S. 1998, \mnras, 297, 251
% high-z galaxy clustering

\bibitem[Baugh et al.(1998)]{baugh98}
Baugh, C. M., Cole, S., Frenk, C. S. \& Lacey, C. G. 1998, \apj, 498, 504

\bibitem[Benson et al.(2001)]{benson01}
Benson, A.~J., Frenk, C.~S., Baugh, C.~M., Cole, S., \& Lacey, C.~G.\ 2001,
\mnras, 327, 1041
% clustering evolution

\bibitem[Berlind \& Weinberg(2002)]{berlind02}
Berlind, A. A. \& Weinberg, D.H. 2002, \apj, 575, 587

\bibitem[Bernardi et al.(2003)]{bernardi03}
Bernardi, M., et al.\ 2003, \aj, 125, 32
% feature in optical depth at z~3.2

\bibitem[Bi \& Davidsen(1997)]{bi97}
Bi, H. \& Davidsen, A.F. 1997,\apj, 479, 523

\bibitem[Bullock et al.(2001)]{bullock01}
Bullock, J.S., Wechsler, R.H., \& Somerville, R.S. 2001, \mnras, 329, 246B

\bibitem[Cen(1992)]{cen92}
Cen, R. 1992, \apjs, 78, 341
% methods paper

\bibitem[Cen(1997)]{cen97}
Cen, R.\ 1997, \apjl, 479, L85
% high end of flux PDF as test of fluctuation amplitude

\bibitem[Cen et al.(1994)]{cen94}
Cen, R., Miralda-Escud\'e, J., Ostriker, J.P., \& Rauch, M. 1994,
\apj, 437, L9
% Ly-alpha forest from gravitational collapse

\bibitem[Cen \& Ostriker(2000)]{cen00}
Cen, R. \& Ostriker, J. P. 2000, \apj, 538, 83

\bibitem[Chen et al.(2002)]{chen02}
Chen, X., Weinberg, D.~H., Katz, N., \& Dave', R. 2002, \apj, submitted,
astro-ph/0203319
% X-ray forest

\bibitem[Cooray(2002)]{cooray02a}
Cooray, A. 2002, \apj, 576, L105
% <N(N-1)> from PSCz

\bibitem[Cooray \& Sheth(2002)]{cooray02b}
Cooray, A. \& Sheth, R. 2002, Phys Rep, 372, 1
% halo model review article

\bibitem[Croft et al.(1997)]{croft97}
Croft, R.A.C., Weinberg, D.H., Katz, N., Hernquist, L., 1997, \apj, 488, 532

\bibitem[Croft et al.(1998)]{croft98}
Croft, R. A. C., Weinberg, D.H., Katz, N., \& Hernquist, L. 1998,
\apj, 495, 44

\bibitem[Croft et al.(1999)]{croft99}
Croft, R. A. C., Weinberg, D.H., Pettini, M., Katz, N., \& Hernquist, L. 1999,
\apj, 520, 1

\bibitem[Croft et al.(2001)]{croft01}
Croft, R. A. C., Di Matteo, T., Dav\'e, R., Hernquist, L., Katz, N.,
Fardal, M., \& Weinberg, D. H. 2001, \apj, 557, 67
% X-ray background

\bibitem[Croft et al.(2002a)]{croft02a}
Croft, R.A.C., Hernquist, L., Springel, V., Westover, M., \& White, M. 2002a,
\apj, 580, 634

\bibitem[Croft et al.(2002b)]{croft02b}
Croft, R. A. C., Weinberg, D. H., Bolte, M., Burles, S.,  Hernquist, L., Katz,
N., Kirkman, D., Tytler, D. 2002b, \apj, 581, 20

\bibitem[Dav\'e et al.(1997)]{dave97}
Dav\'e, R., Dubinski, J., \& Hernquist, L. 1997, New Astronomy, 2, 277

\bibitem[Dav\'e et al.(1999)]{dave99}
Dav\'e, R., Hernquist, L., Katz, N., \& Weinberg, D.H. 1999, \apj, 511, 521

\bibitem[Dav\'e et al.(2001)]{dave01}
Dav\'e, R., Cen, R., Ostriker, J. P., Bryan, G. L., Hernquist, L., Katz, N.,
Weinberg, D. H., Norman, M. L., \& O'Shea, B. 2000, \apj, 552, 473
% WHIM

\bibitem[Dav\'e et al.(2002)]{dave02}
Dav\'e, R., Katz, N., \& Weinberg, D. H. 2002, \apj, 579, 23
% X-ray scaling relations

\bibitem[Fardal et al.(2002)]{fardal02}
Fardal, M.\ A., Katz, N., Weinberg, D. H., Dav\'e, R., \& Hernquist, L. 2002,
\apj, in press, astro-ph/0107290
% sub-mm emission

\bibitem[Fardal et al.(2001)]{fardal01}
Fardal, M.\ A., Katz, N., Gardner, J.P., Hernquist, L., Weinberg, D. H., \& Dav\'e, R. 2001, \apj, 562, 605

\bibitem[Fardal \& Shull(1993)]{fardal93}
Fardal, M.~A.~\& Shull, J.~M.\ 1993, \apj, 415, 524
% correlations of random Lya clouds, UVB fluctuations

\bibitem[Gelb \& Bertschinger(1994)]{gelb94}
Gelb, J. M., \& Bertschinger, E. 1994, \apj, 436, 467
% CDM 1: CDM 1: formation of dark halos

%\bibitem[Giavalisco et al.(1998)]{giavalisco98}
%Giavalisco, M., Steidel, C.C., Adelberger, K.L., Dickinson, M.E., Pettini, M.
%\& Kellogg, M. 1998, \apj, 503, 543

\bibitem[Giavalisco \& Dickinson(2001)]{giavalisco01}
Giavalisco, M.~\& Dickinson, M.\ 2001, \apj, 550, 177
% clustering segregation of LBGs

\bibitem[Gnedin \& Hui(1998)]{gnedin98}
Gnedin, N. Y., \& Hui, L. 1998, \mnras, 296, 44
% HPM technique

\bibitem[Governato et al.(1998)]{governato98}
Governato, F., Baugh, C.M., Frenk, C.S., Cole, S., Lacey, C.G., Quinn, T.R. \&
Stadel, J. 1998, Nature, 392, 359
% semi-analytic + N-body clustering of high-z galaxies

\bibitem[Haardt \& Madau(1996)]{haardt96}
Haardt, F. \& Madau, P. 1996, \apj, 461, 20

\bibitem[Hernquist \& Katz(1989)]{hernquist89}
Hernquist, L. \& Katz, N. 1989, \apjs, 70, 419

\bibitem[Hernquist et al.(1996)]{hernquist96}
Hernquist, L., Katz, N., Weinberg D.H. \& Miralda-Escud\'e 1996 \apjl, 457, L51

\bibitem[Hui \& Gnedin(1997)]{hui97}
Hui, L. \& Gnedin, N.Y. 1997, \mnras, 292, 27

\bibitem[Jing, Mo, \& B\"orner(1998)]{jing98}
Jing, Y. P., Mo, H. J., \& B\"orner, G. 1998, ApJ, 494, 1
% correlation function and pairwise dispersion: CDM vs. LCRS

\bibitem[Kaiser(1984)]{kaiser84}
Kaiser, N. 1984, \apjl, 284, L9

\bibitem[Kaiser(1987)]{kaiser87}
Kaiser, N. 1987, \mnras, 227, 1
% redshift space distortions

\bibitem[Katz, Weinberg \& Hernquist(1996)]{katz96}
Katz, N., Weinberg D.H. \& Hernquist, L. 1996, \apjs, 105, 19 (KWH)

\bibitem[Katz, Hernquist, \& Weinberg(1999)]{katz99}
Katz, N., Hernquist, L., \& Weinberg, D. H. 1999, \apj, 523, 463

\bibitem[Kauffman et al.(1999)]{kauffmann99}
Kauffmann, G., Colberg, J. M., Diaferio, A., \& White, S. D. M. 1999,
\mnras, 307, 529
% N-body + semi-analytic: evolution of clustering

\bibitem[Kollmeier et al.(2002)]{kollmeier02}
Kollmeier, J.A., Weinberg, D.H., Dav\'e, R., \& Katz, N. 2003, in ``The Emergence of Cosmic Structure'' eds. S. Holt \& C. Reynolds, AIP Conference Proceedings, New York, p.191, astro-ph/0212355
% maryland procs

\bibitem[Ma \& Fry(2000)]{ma00}
Ma, C., \& Fry, J. N. 2000, \apj, 543, 503
% halo occupation calculation of P(k) and bispectrum

\bibitem[Madau, Haardt, \& Rees(1999)]{madau99}
Madau, P., Haardt, F., \& Rees, M.~J.\ 1999, \apj, 514, 648
% radiative transfer in clumpy universe, nature of ionizing sources

\bibitem[Marinoni \& Hudson(2002)]{marinoni02}
Marinoni, C., \& Hudson, M. J. 2002, \apj, 569, 101
% HOD from group multiplicity function

\bibitem[McDonald et al.(2000)]{mcdonald00}
McDonald, P., Miralda-Escud\'e, J., Rauch, M., Sargent, W. L. W.,
Barlow, T. A., Cen, R., \& Ostriker, J. P. 2000, \apj, 543, 1
% PDF, power spectrum, and correlation function of Lya forest flux

\bibitem[McDonald et al.(2002)]{mcdonald02}
McDonald, P., Miralda-Escud\'e, J., \& Cen, R. 2002, \apj, 580,42

\bibitem[Miralda-Escud\'e et al.(1996)]{miralda96}
Miralda-Escud\'e J., Cen R., Ostriker, J.P., \& Rauch, M. 1996, \apj, 471, 582
% Lyman-alpha forest tome

\bibitem[Miralda-Escud\'e et al.(1997)]{miralda97}
Miralda-Escud\'e, J., Rauch,. M., Sargent, W.L.W.,
Barlow, T.A., Weinberg, D.H., Hernquist, L., Katz, N., Cen, R. \&
Ostriker, J.P. 1998, in Proc. of the 13th IAP Colloquium,
Structure and Evolution of the IGM
from QSO Absorption Line Systems, eds. P. Petitjean \& S. Charlot,
(Paris: Nouvelles Fronti\`eres), p. 155, astro-ph/9710230
% joint probability distribution of flux decrement

\bibitem[Mo \& Fukugita(1996)]{mof96}
Mo, H.J. \& Fukugita, M. 1996, \apjl, 467, L9
% nature of Lyman break galaxies

\bibitem[Mo \& White(1996)]{mow96}
Mo, H.J., \& White S.D.M. 1996, \mnras, 282, 1096
% analytic halo clustering

\bibitem[Moscardini et al.(1998)]{moscardini98}
Moscardini, L., Coles, P., Lucchin, F. \& Matarrese, S. 1998,
\mnras, 299, 95
% high-z clustering

\bibitem[Moustakas \& Somerville(2002)]{moustakas02}
Moustakas, L.~A.~\& Somerville, R.~S.\ 2002, \apj, in press, astro-ph/0110548
% HOD analysis of EROs, LBGs, and BCEs

\bibitem[Murali et al.(2002)]{murali02}
Murali, C., Katz, N., Hernquist, L., Weinberg, D. H., \& Dav\'e, R. 2002,
\apj, 571, 1
% growth of galaxies

\bibitem[Nagamine et al.(2001)]{nagamine01}
Nagamine, K., Fukugita, M., Cen, R., Ostriker, J.P., 2001, \mnras, 327L, 10N

\bibitem[Peacock \& Smith(2000)]{peacock00}
Peacock, J. A., \& Smith, R. E. 2000, \mnras, 318, 1144
% halo occupation numbers and bias

\bibitem[Pearce et al.(2001)]{pearce01}
Pearce, F. R., Jenkins, A., Frenk, C. S.,
White, S. D. M., Thomas, P. A., Couchman, H. M. P., Peacock, J. A., \&
Efstathiou, G. 2001, \mnras, 326, 649

\bibitem[Pettini et al.(2001)]{pettini01}
Pettini, M., Shapley,
A.~E., Steidel, C.~C., Cuby, J., Dickinson, M., Moorwood, A.~F.~M.,
Adelberger, K.~L., \& Giavalisco, M.\ 2001, \apj, 554, 981
% rest-frame optical spectra of LBGs

\bibitem[Pettini et al.(2002)]{pettini02}
Pettini, M., Rix,
S.~A., Steidel, C.~C., Adelberger, K.~L., Hunt, M.~P., \& Shapley, A.~E.\
2002, \apj, 569, 742
% ISM of cB58

\bibitem[Press, Rybicki, \& Schneider(1993)]{press93}
Press, W.H., Rybicki, G.B. \& Schneider, D.P. 1993, \apj, 414, 64

\bibitem[Rauch et al.(1997)]{rauch97}
Rauch, M., Miralda-Escud\'e, J., Sargent, W. L. W., Barlow, T. A.,
Weinberg, D. H., Hernquist, L., Katz, N., Cen, R., \& Ostriker, J. P.,
1997, \apj, 489, 7
% Omega_b from flux decrement distribution, HIRES vs. simulations

\bibitem[Sargent \& Turner(1977)]{sargent77}
Sargent, W. W., \& Turner, E. L. 1977, ApJ, 212, L3
% redshift space distortion

\bibitem[Schirber \& Bullock(2003)]{schirber03}
Schirber, M., \& Bullock, J. S.\ 2003, \apj, 584, 110
% faint AGN and UV background

\bibitem[Scoccimarro et al.(2001)]{scoccimarro01}
Scoccimarro, R., Sheth, R. K., Hui, L., \& Jain, B. 2001, \apj, 546, 20
% halo occupation distribution

\bibitem[Scott et al.(2000)]{scott00}
Scott, J., Bechtold, J., Dobrzycki,
A., \& Kulkarni, V.~P.\ 2000, \apjs, 130, 67
% UVB from proximity effect at high z

\bibitem[Scranton(2002)]{scranton02}
Scranton, R. 2002, \mnras, 332, 697
% red and blue galaxies in halo model

\bibitem[Seljak(2000)]{seljak00}
Seljak, U. 2000, \mnras, 318, 203
% halo occupation model for bias and clustering

\bibitem[Sheth, Mo, \& Tormen(2001)]{sheth01}
Sheth, R.~K., Mo, H.~J., \& Tormen, G.\ 2001, \mnras, 323, 1
% mass function and biasing with ellipsoidal collapse

\bibitem[Steidel et al.(1996)]{steidel96}
Steidel, C. C., Giavalisco, M., Pettini, M., Dickinson, M.,
\& Adelberger, K. L. 1996, \apj, 462, L17
% spectroscopic confirmation of a population of star-forming galaxies at z>3

\bibitem[Steidel et al.(2000)]{steidel00}
Steidel, C.C., Adelberger, K.L, Shapley, A., Pettini, M., Dickinson, M.,
\& Giavalisco, M. .\ 2000, \apj, 532, 170

\bibitem[Steidel et al.(2002)]{steidel02} 
Steidel, C., Hunt, M.,Shapley, A., Adelberger, K., Pettini, M.,
Dickinson, M., \& Giavalisco, M.\ 2002, \apj, 576, 653
% AGN in LBGs

\bibitem[Steidel, Pettini, \& Adelberger(2001)]{steidel01}
Steidel, C.C., Pettini, M., \& Adelberger, K.~L.\ 2001, \apj, 546, 665
% Lyman continuum radiation from galaxies at z~3.4

\bibitem[Theuns et al.(1998)]{theuns98}
Theuns, T., Leonard, A.,Efstathiou, G., Pearce, F. R., \& Thomas,
P. A. 1998, \mnras, 301, 478
% P3MSPH simulations of lya forest

\bibitem[Weinberg et al.(1999a)]{weinberg99a}
Weinberg, D. H., et al.\ 1999a, in Evolution of Large Scale Structure:
From Recombination to Garching, eds. A.J. Banday, R. K. Sheth,
\& L. N. Da Costa, (Twin Press: Vledder NL), p. 346, astro-ph/9810142

\bibitem[Weinberg et al.(1999b)]{weinberg99b}
Weinberg, D. H., Dav\'e, R., Gardner, J. P., Hernquist, L., \& Katz, N. 1999b,
in Photometric Redshifts and High Redshift Galaxies, eds. R. Weymann,
L. Storrie-Lombard, M. Sawicki, \& R. Brunner,
ASP Conference Series, San Francisco, p. 341, astro-ph/9908133

\bibitem[Weinberg, Hernquist \& Katz(1997)]{weinberg97a}
Weinberg, D.H., Hernquist, L. \& Katz, N. 1997, \apj, 477, 8

\bibitem[Weinberg, Katz, \& Hernquist(1998)]{weinberg98}
Weinberg, D. H., Katz, N., \& Hernquist, L. 1998,
in Origins, eds. J. M. Shull, C. E. Woodward, \& H. Thronson,
(ASP Conference Series: San Francisco), p. 21, astro-ph/9708213

\bibitem[Weinberg et al.(2002a)]{weinberg02a}
Weinberg, D.H., Dav\'e, R., Katz, N., \& Hernquist, L. 2002a, in preparation
% L144 clustering paper

\bibitem[Weinberg et al.(2002b)]{weinberg02b}
Weinberg, D.H., Hernquist, L., Katz, N., 2002b, \apj, 571, 15
% high-z galaxies paper

\bibitem[Weinberg et al.(2003)]{weinberg03}
Weinberg, D.H., Dav\'e, R., Katz, N., \& Kollmeier, J.A. 2003, in ``The Emergence of Cosmic Structure'' eds. S. Holt \& C. Reynolds, AIP Conference Proceedings, New York, p.157, astro-ph/0301186
% maryland procs


\bibitem[Yoshikawa et al.(2001)]{yoshikawa01}
Yoshikawa, K., Taruya, A., Jing, Y. P., Suto, Y. 2001, \apj, 558, 520
% biasing in hydro simulations

\bibitem[Zehavi et al.(2003)]{zehavi03} 
Zehavi, I., Weinberg, D.H., Zheng, Z., Berlind, A.A., Frieman, J.A.,
Scoccimarro, R., Sheth, R.K., Blanton, M.R., Tegmark, M., Mo, H.J., et
al. 2003, astro-ph/0301280

\bibitem[Zhang, Anninos, \& Norman(1995)]{zhang95}
Zhang, Y., Anninos, P., \& Norman, M.L. 1995, \apj, 453, L57
% first Illinois ly-a forest paper

\end{thebibliography}
\end{document}